\documentclass[twocolumn]{aastex631}

\begin{document}

\title{The shapes and dynamical states of dark matter halos across the full mass spectrum in $\Lambda$CDM}

\author{Hang Yang}
\affiliation{Institute for Frontiers in Astronomy and Astrophysics, Beijing Normal University, Beijing 102206, China}
\affiliation{School of Physics and Astronomy, Beijing Normal University, Beijing 100875, China}
\affiliation{Institute for Astrophysics, School of Physics, Zhengzhou University, Zhengzhou 450001, China}
\email{hyang@nao.cas.cn, lgao@bnu.edu.cn}

\author{Liang Gao}
\affiliation{Institute for Frontiers in Astronomy and Astrophysics, Beijing Normal University, Beijing 102206, China}
\affiliation{School of Physics and Astronomy, Beijing Normal University, Beijing 100875, China}
\affiliation{Institute for Astrophysics, School of Physics, Zhengzhou University, Zhengzhou 450001, China}

\author{Qi Guo}
\affiliation{Institute for Frontiers in Astronomy and Astrophysics, Beijing Normal University, Beijing 102206, China}
\affiliation{School of Physics and Astronomy, Beijing Normal University, Beijing 100875, China}

\author{Ming Li}
\affiliation{National Astronomical Observatories (NAOC), Chinese Academy of Sciences, Beijing 100101, China}
\affiliation{School of Astronomy and Space Science, University of Chinese Academy of Sciences, Beijing 100048, China}

\author{Yizhou Liu}
\affiliation{Institute for Frontiers in Astronomy and Astrophysics, Beijing Normal University, Beijing 102206, China}
\affiliation{School of Physics and Astronomy, Beijing Normal University, Beijing 100875, China}

\author{Jie Wang}
\affiliation{National Astronomical Observatories (NAOC), Chinese Academy of Sciences, Beijing 100101, China}
\affiliation{School of Astronomy and Space Science, University of Chinese Academy of Sciences, Beijing 100048, China}

\author{Qiao Wang}
\affiliation{National Astronomical Observatories (NAOC), Chinese Academy of Sciences, Beijing 100101, China}
\affiliation{School of Astronomy and Space Science, University of Chinese Academy of Sciences, Beijing 100048, China}

\author{Simon D.M. White}
\affiliation{Max Planck Institute for Astrophysics, Karl-Schwarzschild-Strasse 1, 85740 Garching, Germany}

\begin{abstract}

We combine the VVV and HyperMillennium simulations to investigate the shapes and dynamical states of dark matter halos from $M_{200}\sim10^{15.5}M_{\odot}$ down to $M_{200}\sim10^{-6}M_{\odot}$, approaching the free streaming cutoff scale for WIMP dark matter. We find halo shapes exhibit a remarkably weak dependence on mass below $M_{200}\sim10^{10}M_{\odot}$, forming a broad plateau with median axis ratios of $s=c/a\sim0.74$ and $q=b/a\sim0.87$. Above this mass, halos become progressively less spherical and more prolate, with median axis ratios decreasing to $s\sim0.42$ and $q\sim0.56$ at the high mass end. The median axis ratios are well described by smoothly broken power laws. A similar low mass plateau is also found in the dynamical states of halos. The low mass plateau in halo shape is present in both relaxed and unrelaxed populations, with relaxed halos generally being rounder than unrelaxed halos at fixed mass. The median $s$ is systematically smaller in the inner regions than in the outskirts at all masses. Moreover, high mass halos tend to have more prolate interiors, whereas low mass halos tend to have more oblate interiors. At fixed mass, we find that more radially biased velocity anisotropy is associated with smaller axis ratios and more prolate shapes. Halo shapes also show moderate correlations with the concentration proxy $V_{\rm max}/V_{200}$, with higher values associated with rounder shapes, but only weak correlations with halo spin. Our results therefore provide a baseline for halo shapes and dynamical states across full mass spectrum in $\Lambda$CDM.

\end{abstract}


\section{Introduction} \label{sec:intro}

The standard $\Lambda$CDM model predicts that dark matter collapses under gravity to form bound structures known as halos. Although their spherically averaged density profiles are well described by the NFW and Einasto models \cite[e.g.][]{1965TrAlm...5...87E, 1996ApJ...462..563N, 2008MNRAS.387..536G}, halos generally exhibit triaxial shapes in three dimensions \cite[e.g.][]{1988ApJ...327..507F, 2002ApJ...574..538J, 2006MNRAS.367.1781A}.

Observations have revealed the departures from circular symmetry in the projected mass distributions and in maps of the hot gas \citep[e.g.][]{2010MNRAS.405.2215O, 2018ApJ...860..104U, 2026A&A...710A..40G}. The mass distribution could be tracked by strong or weak lensing \citep[e.g.][]{2007ApJ...668..643L, 2009ApJ...695.1446E, 2020MNRAS.496.2591O, 2026ApJ..1011..154C}. The configuration of intracluster light (ICL) and satellite galaxies can also be used to probe the halo shapes \citep[e.g.][]{2018MNRAS.475.2421S, 2019MNRAS.482.2838M, 2021MNRAS.504.6033S, 2022MNRAS.514..390G}. In addition, dynamical methods provide complementary constraints on the halo shapes \citep[e.g.][]{2015ApJ...810...22H, 2016ApJ...833...31B, 2024A&A...691A.277W}.

The triaxial shapes of halos can arise at the onset of formation, by the non-spherical density peaks and the surrounding tidal field \citep[e.g.][]{1986ApJ...304...15B, 2001MNRAS.323....1S, 2023MNRAS.518.3509D}. The radial orbit instability can also generate triaxiality during collapse \citep[e.g.][]{2011MNRAS.414.3044V}. Subsequent accretion and mergers can reshape this anisotropy \citep[e.g.][]{VeraCiro2011, 2025ApJ...979..223W}. These processes also influence the internal dynamical states of halos. As halos experience different assembly histories, even halos of similar mass can exhibit markedly different shapes and dynamical states \citep[e.g.][]{2020ApJ...899...81C, 2021MNRAS.500.1029L}. Despite this diversity, an important question is whether halos exhibit systematic statistical trends with mass and how halo shapes are correlated with their dynamical states.

Previous simulations have shown that more massive halos tend to have smaller axis ratios and this mass dependence is often described by a single power law over the mass ranges probed \citep[e.g.][]{2002ApJ...574..538J, 2006MNRAS.367.1781A, Bett2007, Henson2017, Chua2019}. Halo shapes also vary with radius, with massive halos tend to be more prolate in their inner regions than in their outskirts \citep[e.g.][]{2007MNRAS.377...50H, VeraCiro2011, Chua2019}. Halo dynamical states also depend on mass, with more massive halos generally being less dynamically relaxed \citep[e.g.][]{Neto2007, Power2012}. However, previous studies have primarily focused on galaxy  and cluster mass halos, and a systematic investigation of halo shapes and dynamical states across a broad mass range, particularly extending to much smaller halos, remains lacking.

Here we combine the VVV and HyperMillennium (HM) simulations to investigate the shapes and dynamical states of dark matter halos at $z=0$ across more than twenty orders of magnitude in mass. The nested zooms of VVV resolve the internal structure of planetary mass halos with $M_{200}\sim10^{-6}\,M_\odot$ \citep{2020Natur.585...39W}. At the opposite end, HM follows $16128^3$ dark matter particles in a volume 125 times that of the VVV parent box, allowing us to sample rare massive clusters with high resolution \citep{2026MNRAS.548ag540W}. Together, these simulations provide an unprecedented dynamic range for a systematic study of the halo shapes and dynamical properties across the full halo mass spectrum in a cosmological context. Our results provide a comprehensive statistical characterization of halo shapes and their connections to internal dynamical states, establishing a reference for understanding halo structure across the full mass spectrum in the standard $\Lambda$CDM universe.

The structure of this paper is as follows. In Section 2 we present the numerical simulations used in this work and the data analysis methods. Section 3 describes the results on halo shapes and dynamical states. In Section 4, we summarise and discuss our results.

\section{Sample and Analysis Procedure}


\begin{table*}
\centering
\caption{Basic parameters of the HM and VVV simulations. The second column
gives the box side length for HyperMillennium and L0, and the radius of the
high-resolution region for the zoom simulations. Here $\epsilon$ is the
gravitational softening length and $m_{\rm p}$ is the particle mass.
The final column gives the mean density of each region at $z=0$ in units
of the cosmic mean.}
\label{tab: the parameters of simulation}
\begin{tabular}{lrrrr}
\toprule
Simulation &
$L_{\rm box}$ [$h^{-1}\,\mathrm{Mpc}$] &
$\epsilon$ [$h^{-1}\,\mathrm{kpc}$] &
$m_{\rm p}$ [$h^{-1}M_\odot$] &
$\bar{\rho}/\rho_{\rm mean}$ \\
\hline
Hyper & 2500    & 3.0                & $3.2\times10^{8}$   & 1.0   \\
L0    & 500     & 5.0                & $1.0\times10^{9}$   & 1.0   \\
L1    & 35      & $3.0\times10^{-1}$ & $5.0\times10^{5}$   & 0.39  \\
L2    & 6.0     & $3.8\times10^{-2}$ & $1.0\times10^{3}$   & 0.082 \\
L3    & 0.68    & $5.6\times10^{-3}$ & 1.9                 & 0.036 \\
L4    & 0.18    & $6.8\times10^{-4}$ & $3.7\times10^{-3}$  & 0.026 \\
L5    & 0.024   & $1.5\times10^{-4}$ & $3.9\times10^{-5}$  & 0.024 \\
L6    & 0.0045  & $2.6\times10^{-5}$ & $1.8\times10^{-7}$  & 0.014 \\
L7    & 0.00075 & $3.6\times10^{-6}$ & $5.8\times10^{-10}$ & 0.016 \\
\hline
\end{tabular}
\end{table*}

\subsection{Simulations}

In this paper, we combine the VVV simulations
\citep{2020Natur.585...39W} with the HyperMillennium (HM) simulation \citep{2026MNRAS.548ag540W} to investigate dark matter halo properties across more than twenty orders of magnitude in mass, from planetary mass halos to the most massive galaxy clusters halos at present day.

The VVV simulations were performed with \textsc{Gadget-4} \citep{2021MNRAS.506.2871S} and employ a sequence of nested cosmological zoom-in initial conditions, starting from a periodic parent volume of side length $500 h^{-1}\mathrm{Mpc}$. At each successive zoom-in level, a nearly spherical underdense region is selected from the parent simulation and resimulated at higher resolution. In this work, we use levels L0–L6 and the L7c run, the latter incorporating a small-scale cutoff in the initial power spectrum corresponding to the free-streaming suppression induced by a thermal weakly interacting massive particle (WIMP) with a mass of approximately $100,\mathrm{GeV}$.

At the highest cluster mass end, we adopted the HM simulation. The HM simulation is performed with the \textsc{PhotoNs} \citep{2018RAA....18...62W, 2021RAA....21....3W}, which follows $16128^{3}$ dark matter particles (about 4.2 trillion) in a periodic comoving box of side length $2.5\,h^{-1}\,\mathrm{Gpc}$, with a particle mass of $3.2\times10^{8}\,h^{-1}\,\mathrm{M}_{\odot}$ and a gravitational softening length of $3.0\,h^{-1}\,\mathrm{kpc}$. 

Both simulation suites adopt closely flat $\Lambda$CDM cosmologies. The VVV simulations use $\Omega_{\rm m}=0.307$, $\Omega_{\Lambda}=0.693$, $h=0.6777$,
$n_{\rm s}=0.9611$, and $\sigma_{8}=0.8288$, following
\cite{2014A&A...571A..16P}. HM adopts
$\Omega_{\rm m}=0.3111$, $\Omega_{\Lambda}=0.6889$,
$\Omega_{\rm b}=0.049$, $h=0.6766$, $n_{\rm s}=0.9665$,
and $\sigma_{8}=0.8102$, following \cite{2020A&A...641A...6P}. The cosmological parameters common to these two sets differ by less than $3$ per cent.

Dark matter haloes were identified using the standard friends-of-friends (FOF) and SUBFIND in both simulations. In Table~\ref{tab: the parameters of simulation}, we list the basic parameters of each run being used in this work.

\subsection{Halo sample selection and mass binning}

\begin{table*}
\centering
\small
\caption{Mass bins retained in this work. Mass intervals and median values are given in $\log_{10}(M_{200}/M_{\odot})$. The median logarithmic mass in each bin is used as its plotting coordinate. Overlapping mass intervals from different simulation levels are listed separately. The minimum particle number cut lead to lower halo counts in some low mass bins than in adjacent higher mass bins.}
\label{tab:review_mass_bins}
\setlength{\tabcolsep}{4pt}
\begin{tabular}{@{}lrrr@{\hspace{2em}}lrrr@{}}
\hline\hline
Sample & Median & Mass interval & $N_{\rm halo}$ &
Sample & Median & Mass interval & $N_{\rm halo}$ \\
\hline
L7 & $-5.4$ & $[-6,-5)$ & 1372 & L2 & $6.8$  & $[6,7)$   & 5382 \\
L7 & $-4.7$ & $[-5,-4)$ & 725  & L2 & $7.3$  & $[7,8)$   & 3617 \\
L7 & $-3.7$ & $[-4,-3)$ & 101  & L2 & $8.3$  & $[8,9)$   & 351 \\
L6 & $-3.1$ & $[-4,-3)$ & 258  & L2 & $9.2$  & $[9,10)$  & 30 \\
L6 & $-2.8$ & $[-3,-2)$ & 725  & L1 & $9.5$  & $[9,10)$  & 22847 \\
L6 & $-1.8$ & $[-2,-1)$ & 41   & L1 & $10.3$ & $[10,11)$ & 6226 \\
L5 & $-0.5$ & $[-1,0)$  & 2375 & L1 & $11.3$ & $[11,12)$ & 721 \\
L5 & $0.2$  & $[0,1)$   & 422  & L1 & $12.3$ & $[12,13)$ & 65 \\
L5 & $1.2$  & $[1,2)$   & 34   & L0 & $12.9$ & $[12.5,13.5)$ & 132983 \\
L4 & $1.4$  & $[1,2)$   & 3104 & L0 & $13.7$ & $[13.5,14.5)$ & 16348 \\
L4 & $2.3$  & $[2,3)$   & 509  & L0 & $14.6$ & $[14.5,15.5)$ & 553 \\
L4 & $3.3$  & $[3,4)$   & 57   & Hyper & $15.5$ & $[15.4,15.6)$ & 98 \\
L3 & $3.9$  & $[3,4)$   & 743  & & & & \\
L3 & $4.3$  & $[4,5)$   & 2,983 & & & & \\
L3 & $5.3$  & $[5,6)$   & 260  & & & & \\
\hline
\end{tabular}
\end{table*}

We select central haloes from the VVV simulations at $z=0$ and bin each simulation level independently in intervals of width $1$ dex in $\log_{10} M_{200}$. To extend the sample towards the high mass end, we additionally include central halos from the HM catalogue with masses in the range $10^{15.4}<M_{200}/M_{\odot}<10^{15.6}$. To ensure adequate numerical resolution, we require each halo to contain at least 3000 particles. This resolution cut removes some halos from the low mass bins, resulting in reduced sample sizes in these bins. We use the median mass of the retained haloes as the representative mass for each bin, so the plotting coordinates are determined directly from the selected sample. Table~\ref{tab:review_mass_bins} summarizes the mass intervals, median masses, and halo counts for all samples retained in this work.

\subsection{Halo shapes}

We model dark matter halos as triaxial systems and characterize their shapes using a modified inertia tensor defined as

\begin{equation}
I_{ij} = \frac{\sum_n m_n x_i^n x_j^n / w_n}{\sum_n m_n},
\end{equation}

where $x_i^n$ is the $i$-th component of the position vector of the $n$-th particle relative to the halo centre, defined as the location of the minimum gravitational potential, $m_n$ is its mass, and $w_n$ is a weight factor. Two commonly used choices for the weight factor are $w_n=1$ and $w_n=(r_{\rm ell}^n)^2$. The former defines the standard (or unweighted) inertia tensor, while the latter defines the reduced inertia tensor, where

\begin{equation}    
  r^2_{\rm ell} = (\frac{x}{a})^2  + (\frac{y}{b})^2 + (\frac{z}{c})^2
\end{equation}

Here, $a$, $b$, and $c$ are the semi-axis lengths of the ellipsoid. The standard inertia tensor gives relatively greater weight to particles in the outer regions of a halo, whereas the reduced inertia tensor downweights particles at larger ellipsoidal radii, thereby reducing the influence of the outer regions. Further technical details can be found in \cite{2011ApJS..197...30Z}.


For our fiducial measurements in this work, we determine the local halo shapes within ellipsoidal shells centered at three radii, $r/R_{200}=0.15$, $0.3$, and $1$, each with a width of $0.1R_{200}$. At each radius, we iteratively diagonalize the reduced inertia tensor ($w=r_{\rm ell}^2$). In each iteration, particle coordinates are transformed into the principal axis frame, the axis ratios are updated using the eigenvalues of the inertia tensor, and particles are reselected within the corresponding ellipsoidal shell. This procedure is repeated until the relative changes in the axis ratios are smaller than $1\%$. We then obtain the final inertia tensor $I_{ij}$.

Let $\lambda_1 \geq \lambda_2 \geq \lambda_3$ denote the eigenvalues of $I_{ij}$. We define the minor-to-major and intermediate-to-major axis ratios as $s=c/a=\sqrt{\lambda_3/\lambda_1}$ and $q=b/a=\sqrt{\lambda_2/\lambda_1}$, respectively. The triaxiality parameter is defined as
\begin{equation}
T = \frac{1-q^2}{1-s^2}.
\end{equation}

Our fiducial measurements include all particles surrounding the central halo, including those associated with surviving subhalos. To assess the impact of substructure on the measured halo shapes, we repeat the shell based measurements after excluding particles assigned to subhalos.

We also investigate the dependence of the measured shapes on the adopted measurement definition by comparing the local shell based measurements with global halo shapes. The latter are obtained using an iterative procedure in which the shape of an ellipsoidal aperture is updated, while its semi-major axis is fixed at $R_{200}$. The resulting axis ratios are denoted by $q_{\rm vol}$ and $s_{\rm vol}$, respectively. Comparisons between these different shape measurements are presented in Appendices~\ref{sec:A1} and \ref{sec:A2}.

Finally, the underdense zoom regions in VVV may introduce an environmental selection bias relative to a representative halo population at the same mass. We examine the sensitivity of our results to this effect in Appendix~\ref{sec:A3} and we find no significant dependence of halo shapes on the local environment \citep[see also][]{2007MNRAS.378...55M, 2007MNRAS.375..489H}.

\subsection{Halo dynamical states}

We characterize halo dynamical states and internal properties using the following parameters.

\begin{enumerate}

\item \textit{Centre-of-mass offset, $X_{\rm off}$}.

The centre-of-mass offset quantifies the displacement between the halo centre, defined as the position of the minimum gravitational potential, and the centre of mass of all particles within $R_{200}$. It is normalized by $R_{200}$ and defined as
\begin{equation}
X_{\rm off}=\frac{|{r}_{\rm cm}-{r}_{\rm c}|}{R_{200}},
\end{equation}
where ${r}_{\rm cm}$ and ${r}_{\rm c}$ denote the centre-of-mass position and the position of the potential minimum, respectively.

\item \textit{Virial ratio, $\eta$.}

We define the virial ratio as
\begin{equation}
\eta = \frac{2K}{|U|},
\end{equation}
where $K$ and $U$ are the total kinetic and gravitational potential energies, respectively, of all particles within $R_{200}$. Following \citet{Neto2007}, we estimate the gravitational potential energy using a random sample of 2000 particles within $R_{200}$ and summing the gravitational interaction energies over all distinct particle pairs in the sample.

\item \textit{Surface pressure corrected virial ratio, $\eta_{\rm corr}$.}

When the virial ratio is evaluated within $R_{200}$, the flux of particle momentum across the boundary contributes to the virial balance. This contribution can be accounted for by introducing a surface pressure term, $E_{\rm s}$ \citep{Shaw2006,Power2012}, such that the virial relation becomes
\begin{equation}
    2K+U-E_{\rm s}=0.
\end{equation}
Following \citet{Power2012}, we estimate the boundary pressure using dark matter particles in a spherical shell spanning $0.8R_{200}<r\leq R_{200}$. The shell pressure and the corresponding surface term are estimated as
\begin{equation}
    P_{\rm s} = \frac{1}{3V_{\rm sh}}
    \sum_{i\in{\rm sh}} m_i
    \left|{v}_i-{v}_{\rm cm}\right|^2,
\end{equation}
and
\begin{equation}
    E_{\rm s}\simeq 4\pi r_{\rm med}^{3}P_{\rm s},
\end{equation}
respectively, where $V_{\rm sh}$ is the shell volume, ${v}_{\rm cm}$ is the halo centre-of-mass velocity, and $r_{\rm med}$ is the median radius of the particles in the shell. We then define the surface pressure corrected virial ratio as
\begin{equation}
    \eta_{\rm corr} = \frac{2K-E_{\rm s}}{|U|}.
\end{equation}

\item \textit{Concentration indicator, $V_{\rm max}/V_{200}$.}

In this work, we adopt $V_{\rm max}/V_{200}$ as a proxy for halo concentration. Here, $V_{200}$ is the circular velocity at $R_{200}$, defined as
\begin{equation}
V_{200} = \sqrt{\frac{G M_{200}}{R_{200}}},
\end{equation}
where $M_{200}$ is the mass enclosed within $R_{200}$, and $V_{\rm max}$ is the maximum circular velocity of the halo.

\item \textit{Spin parameter, $\lambda$.}

We adopt the dimensionless spin parameter \citep[e.g.][]{2001ApJ...555..240B, 2023MNRAS.518.5253Y} to quantify the degree of rotation of halos,
\begin{equation}
    \lambda = \frac{|{J}_{200}|}{\sqrt{2}M_{200}V_{200}R_{200}},
\end{equation}
where ${J}_{200}$ is the total angular momentum of all particles within $R_{200}$.

\item \textit{Velocity anisotropy parameter, $\beta$.}

The velocity anisotropy parameter is defined as
\begin{equation}
    \beta = 1-\frac{\sigma_\theta^2+\sigma_\phi^2}{2\sigma_r^2},
\end{equation}
where $\sigma_r$, $\sigma_\theta$, and $\sigma_\phi$ are the velocity dispersions in spherical coordinates. Larger values of $\beta$ indicate more radially biased velocity distributions. We calculate $\beta$ using all particles within $R_{200}$ and $0.3R_{200}$, denoting the corresponding measurements by $\beta_{100}$ and $\beta_{30}$, respectively.

\end{enumerate}

\section{Results} \label{sec:res}
\subsection{Illustrative Examples of Halo Shape}

\begin{figure*}
    \includegraphics[width=0.32\textwidth]{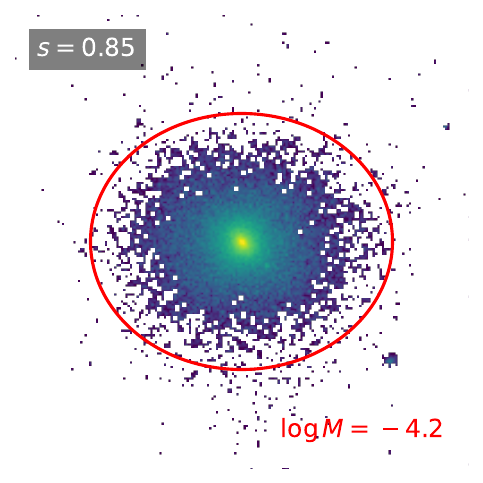}
    \includegraphics[width=0.32\textwidth]{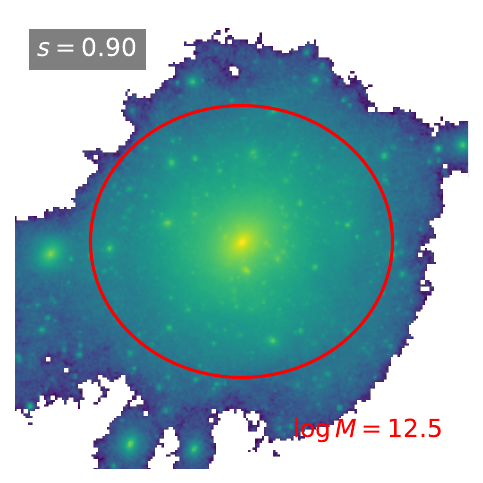}
    \includegraphics[width=0.32\textwidth]{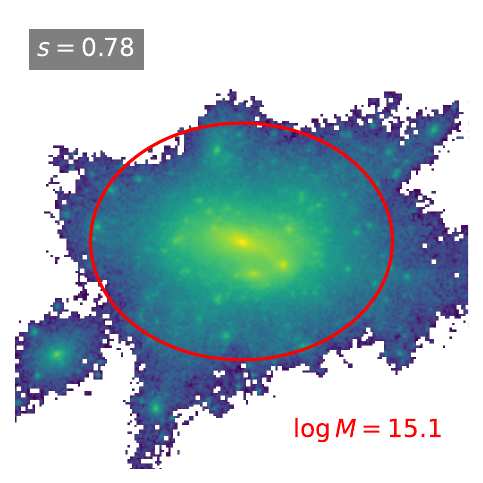}
    \\
     \includegraphics[width=0.32\textwidth]{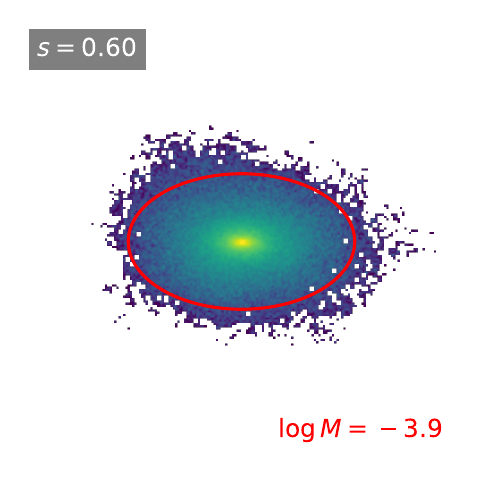}
    \includegraphics[width=0.32\textwidth]{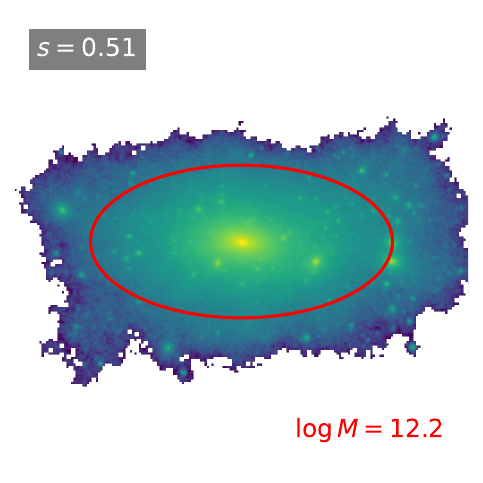}
     \includegraphics[width=0.32\textwidth]{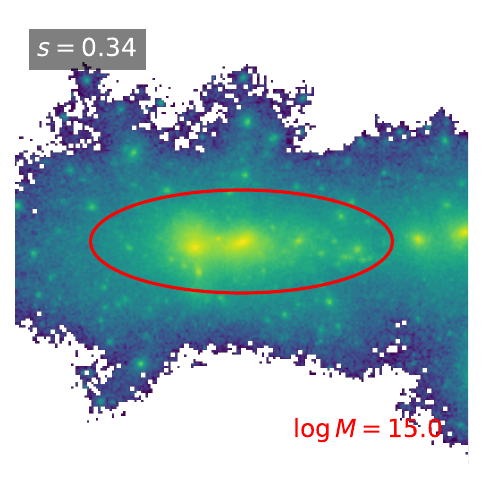}
    \centering
    \caption{Projected dark matter maps of halos at planetary, galactic and cluster masses, from left to right. The upper and lower rows show rounder and more elongated halos, respectively. Each halo is projected along the intermediate axis of its fitted triaxial ellipsoid, so that the image plane is spanned by the major and minor axes. The red lines denote our fitted ellipsoids with major axes equal to the $R_{200}$.}
    \label{fig:halo_examples}
\end{figure*}

Fig.~\ref{fig:halo_examples} illustrates the projected dark matter distributions of planetary, galactic and cluster sized halos at $z=0$. In each mass range, we rank halos by their $s=c/a$ at $R_{200}$ and randomly select one halo from
each of the upper and lower deciles. The red lines denote our fitted ellipsoids with major axes equal to the $R_{200}$. All maps are presented in the coordinate system of the fitted ellipsoid.

The maps illustrate how widely halo shapes can differ at comparable masses. However, even these randomly selected examples hint at a mass trend in halo shape. In the following subsection, we use the full sample to quantify how the median halo shape and its scatter vary with mass.

\subsection{Halo shapes across the full mass spectrum}

\begin{figure}
    \includegraphics[width=\linewidth]{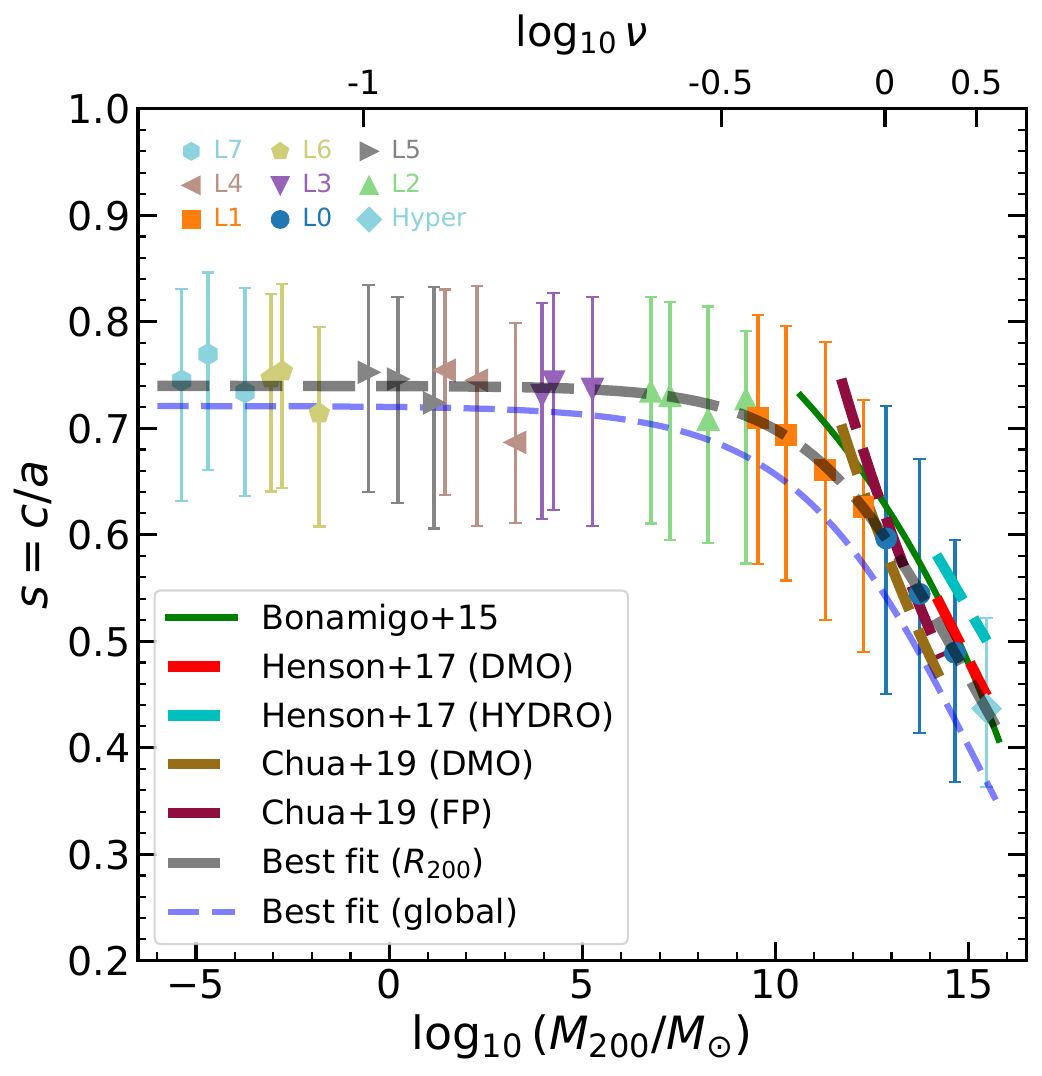}
    \caption{Median DM halo shape parameter, $s=c/a$, as a function of halo mass, measured at $r=R_{200}$. Different colours and symbols denote the VVV resolution levels and the HM sample, as indicated in the legend. Points represent the median values in each mass bin, while error bars span the 16th–84th percentiles of the $s$ distribution. The grey and blue dashed curves show our best fitting relations for the shell and global shape measurements, respectively. The global $s$ values lie slightly below those measured in the outer shell, indicating that halo interiors are generally less spherical than their outskirts. The remaining curves show relations from previous studies, as indicated in the legend. For reference, the corresponding peak height at $z=0$ is shown along the top axis.}
    \label{fig:s_r200}
\end{figure}

We first examine the mass dependence of halo shape, characterized by the axis ratio $s=c/a$, $q=b/a$, and the triaxiality parameter $T=(1-q^2)/(1-s^2)$ at $R_{200}$. Fig.~\ref{fig:s_r200} shows the mass dependence of $s$ across the combined VVV and HM samples. Symbols indicate the median value in each mass bin, with error bars enclosing the 16th--84th percentile range. The upper axis gives the corresponding peak height $\log_{10}\nu$ at $z=0$, where $\nu = \delta_{\rm c}/\sigma(M) \sim 1.68/\sigma(M_{200})$ measures the rarity of density peaks, and $\sigma(M_{200})$ denotes the rms linear density fluctuation on the mass scale $M_{200}$.

The dashed curves show the best fit to the median shape--halo mass relation using a smoothly broken power law,

\begin{equation}
    F(M) = A_0[1+(M/M_t)^{\gamma}]^{-\beta/\gamma}
    \label{eq:shape_fit}
\end{equation}

Here, $A_0$ is the low mass asymptote of the fitted axis ratio, $\beta$ is the high mass power law index. The $M_{\rm t}$ and $\gamma$ control the transition scale and its sharpness, respectively. We fit the median $s$ and $q$ with $F(M)$ and the median $T$ with $1-F(M)$. All best fitting parameters in this work are listed in Table~\ref{tab:fit_params}.

The median sphericity decreases with increasing halo mass, with the most massive cluster halos reaching $s\sim0.43$. Towards lower masses, the relation progressively flattens, approaching an approximately constant value of $s\sim0.74$. This indicates that the strong mass dependence of halo sphericity at galaxy and cluster scales becomes substantially weaker below $M_{200}\sim10^{10}M_{\odot}$. 

\begin{figure*}
    \includegraphics[width=0.5\textwidth]{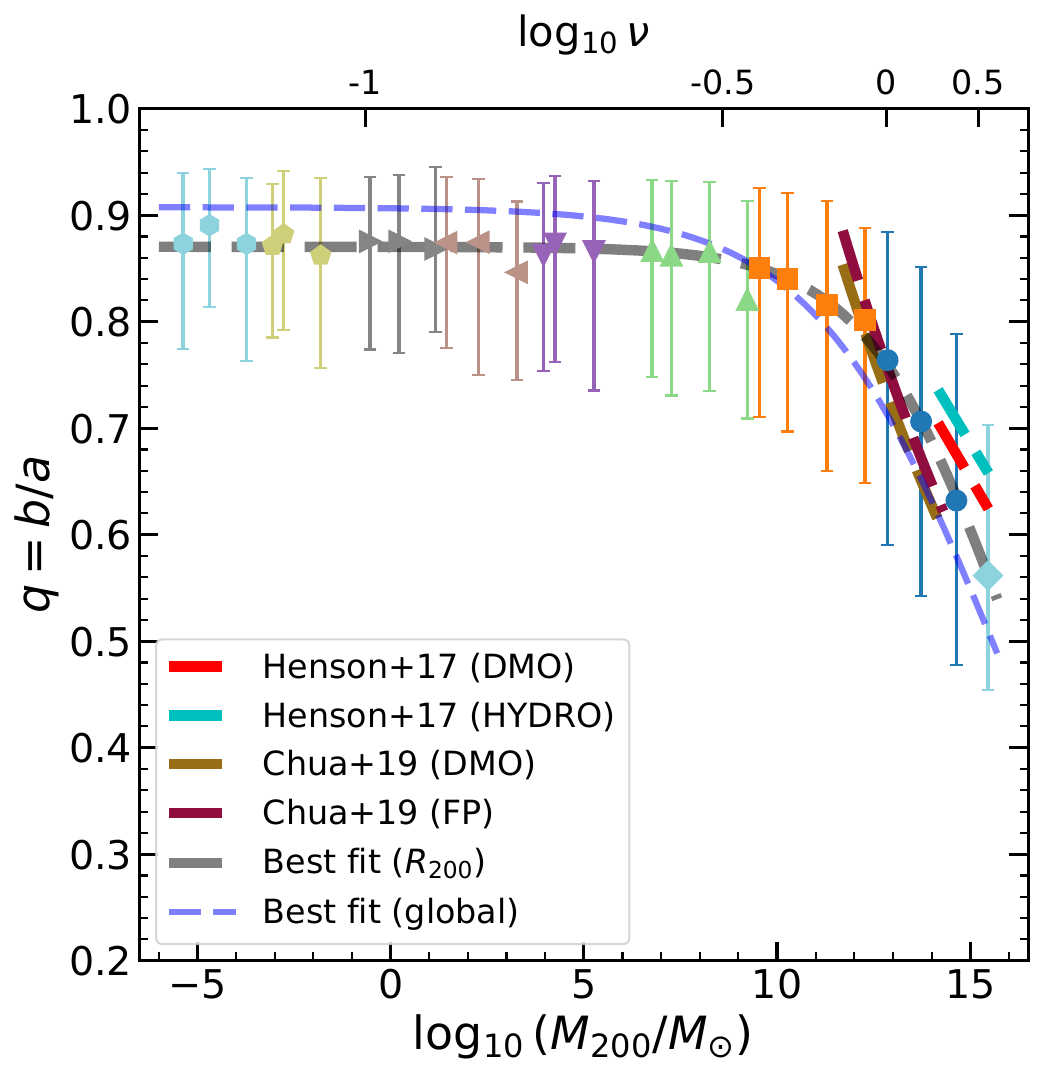}
    \includegraphics[width=0.5\textwidth]{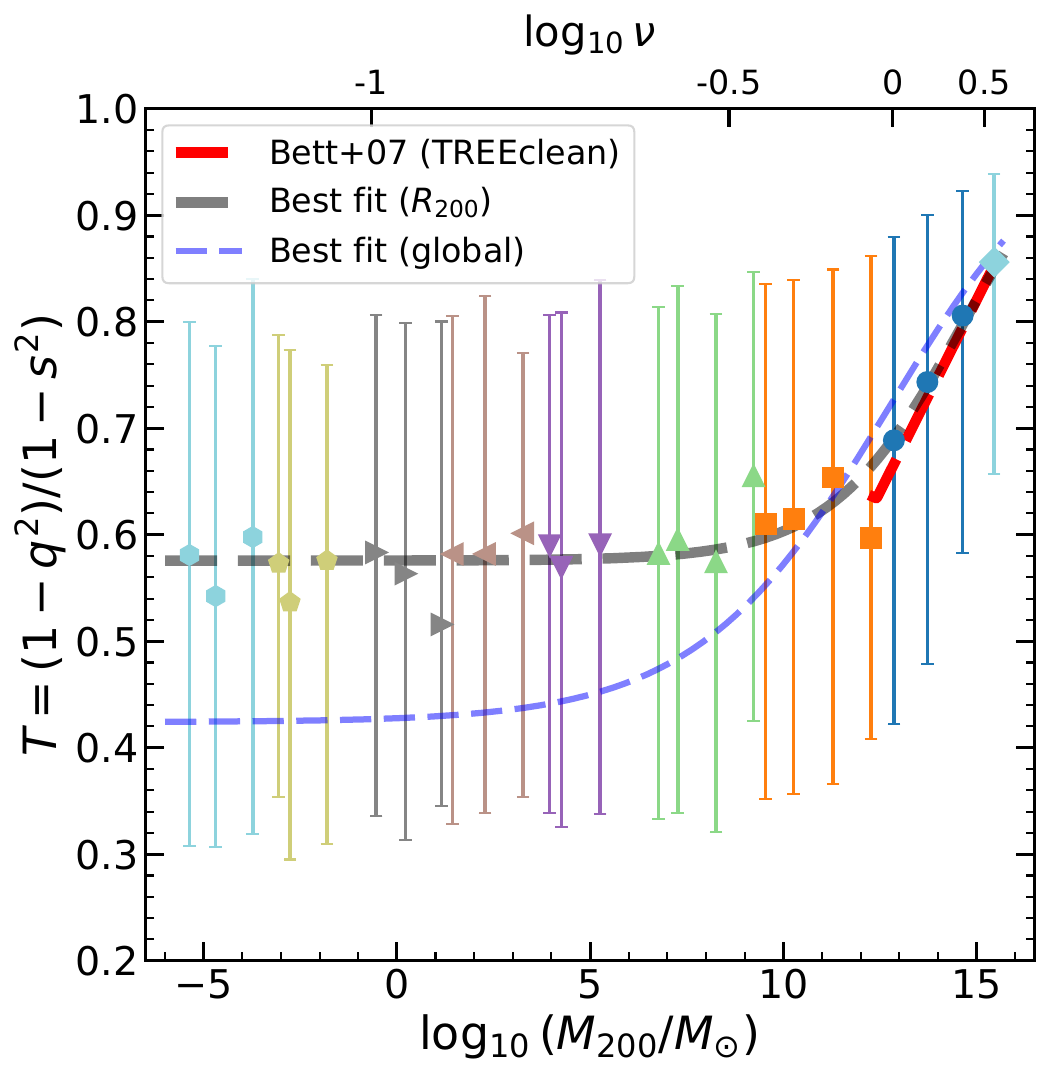}
    \caption{Mass dependence of $q=b/a$ (left) and $T=(1-q^2)/(1-s^2)$ (right), measured at $r=R_{200}$.
    Colours and symbols follow Fig.~\ref{fig:s_r200}. In the right panel, $T$ is evaluated for
    individual halos before binning. The grey and blue dashed curves show our best fitting relations for the shell and global shape measurements, respectively. The differences between the global and outer shell measurements of $q$ and $T$ exhibit a mass dependence, reflecting variations in the internal halo structure across different mass. The remaining curves show relations from previous studies, as indicated in the legend.}
    \label{fig:q_r200}
\end{figure*}

While $s$ quantifies the overall departure from spherical symmetry, $q$ and $T$ provide additional information about the relative axis lengths and the distinction between oblate and prolate configurations. The median $q$ follows a similar mass dependence (Fig.~\ref{fig:q_r200}), remaining near $q\sim0.87$ at low masses before declining to $q\sim0.56$ at the high mass end. The median $T$ remains approximately constant at $T\sim0.57$ over much of the low mass range, before increasing to $T\sim0.86$ in the most massive bin. Together, the decreasing $s$ and increasing $T$ indicate that massive halos are typically less spherical and have increasingly prolate configurations.

In the high mass regime, our shape–mass relations agree with the trends reported in previous studies \citep[e.g.][]{Bett2007, Bonamigo2015, Henson2017, Chua2019}, with our sample extending up to halo masses of $M_{200}\sim10^{15.5}M_{\odot}$. At the low mass end, our results extend these relations down to the cutoff scale, revealing a broad plateau.

The flattening of the median shape–mass relations at low masses does not imply that individual low-mass halos are necessarily close to spherical. As illustrated in Fig.~\ref{fig:halo_examples}, even planet mass halos can exhibit strongly non-spherical morphologies. Thus, the low mass plateau characterises the median population trend rather than the full diversity of individual halo shapes.

For reference, the blue dashed curves in these figures show the best fitting shape–mass relations for the global halo axis ratios $s_{\rm vol}$ and $q_{\rm vol}$. The global and shell based measurements follow similar trends, but differ systematically in their absolute values. This difference reflects variations in halo shape across different radii, which we investigate in detail in the following section.

The published full-physics and DMO results plotted in Fig.~\ref{fig:s_r200} and Fig.~\ref{fig:q_r200} show only modest differences in the outer halo axis ratios. We therefore adopt outer halo shape--mass relations as the fiducial $\Lambda$CDM prediction and expect them to remain a good approximation even when baryonic effects are included.

\begin{table*}
\centering
\small
\caption{Best fitting parameters of the median halo shape--mass relations at different radii. For local shape measurements, the second column gives the central radius of the shell. For $s$ and $q$, $S_{\rm fit}(M)=F(M)$ in equation~\ref{eq:shape_fit}. For the triaxiality parameter $T$, $T_{\rm fit}(M)=1-F(M)$. RMS is the root-mean-square residual of the bin medians.}
\label{tab:fit_params}
\begin{tabular}{lrrrrrr}
\hline\hline
Shape parameter & $r/R_{200}$ & $A_0$ &
$\log_{10}(M_{\rm t}/M_{\odot})$ & $\beta$ & $\gamma$ & RMS \\
\hline
$s_{100}$ & 1    & 0.740 & 13.15 & 0.083 & 0.238 & 0.0147 \\
$s_{30}$  & 0.3  & 0.719 & 12.72 & 0.088 & 0.218 & 0.0128 \\
$s_{15}$  & 0.15 & 0.684 & 12.00 & 0.064 & 0.187 & 0.0235 \\
$q_{100}$ & 1    & 0.870 & 15.0 & 0.129 & 0.246 & 0.0101 \\
$q_{30}$  & 0.3  & 0.914 & 12.08 & 0.066 & 0.242 & 0.0137 \\
$q_{15}$  & 0.15 & 0.896 & 10.05 & 0.044 & 0.243 & 0.0206 \\
$T_{100}$ & 1    & 0.424 & 15.0 & 0.312 & 0.251 & 0.0255 \\
$T_{30}$  & 0.3  & 0.621 & 13.16 & 0.210 & 0.168 & 0.0409 \\
$T_{15}$  & 0.15 & 0.579 & 9.37  & 0.112 & 0.228 & 0.0481 \\
$s_{\rm vol}$ & & 0.721 & 14.55 & 0.140 & 0.191 & 0.0247 \\
$q_{\rm vol}$ & & 0.908 & 15.0 & 0.136 & 0.187 & 0.0174 \\
$T_{\rm vol}$ & & 0.576 & 13.08 & 0.197 & 0.172 & 0.0376 \\
\hline
\end{tabular}
\par\smallskip
\begin{minipage}{0.97\linewidth}
\footnotesize
\end{minipage}
\end{table*}

\subsection{The shapes of inner halos}

\begin{figure*}
    \includegraphics[width=0.32\textwidth]{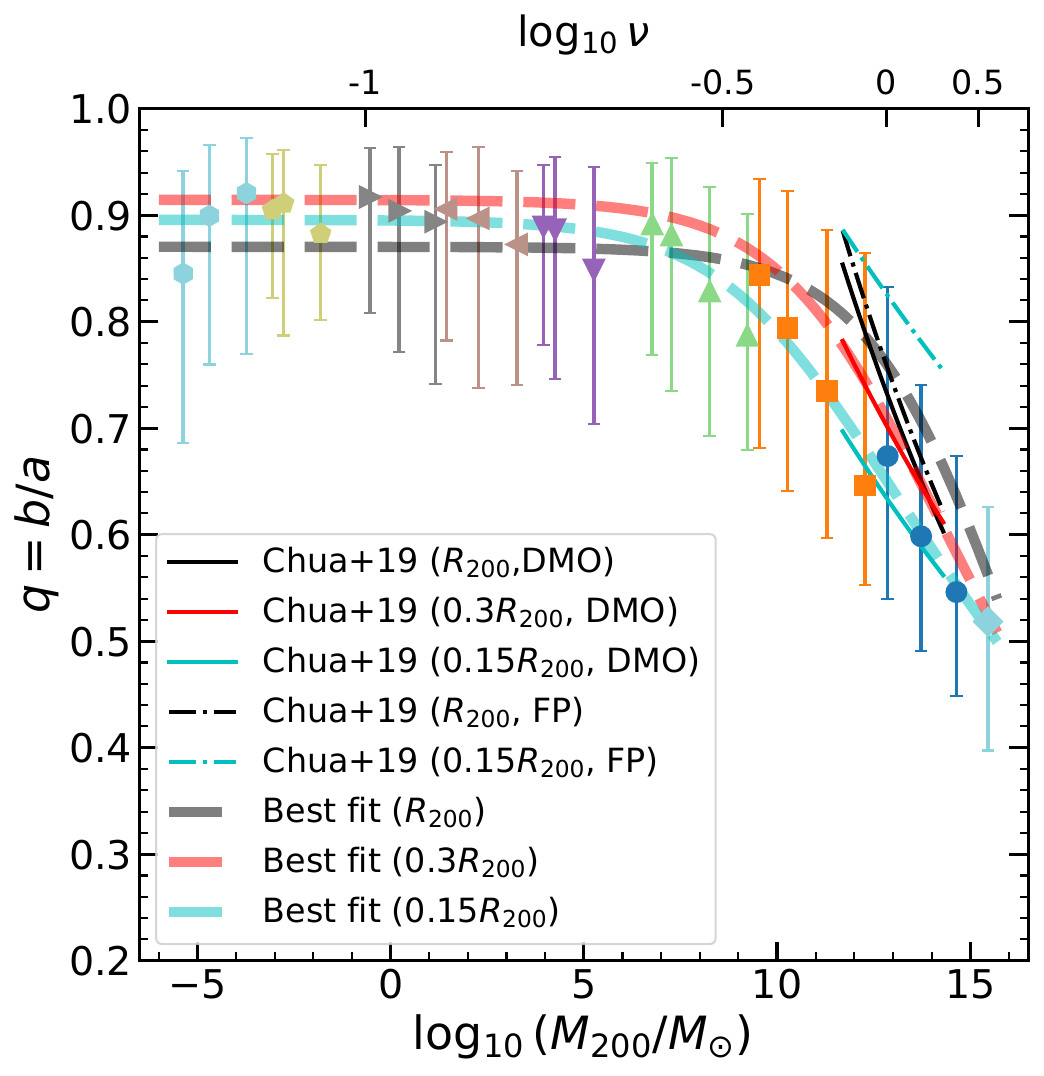}
    \includegraphics[width=0.32\textwidth]{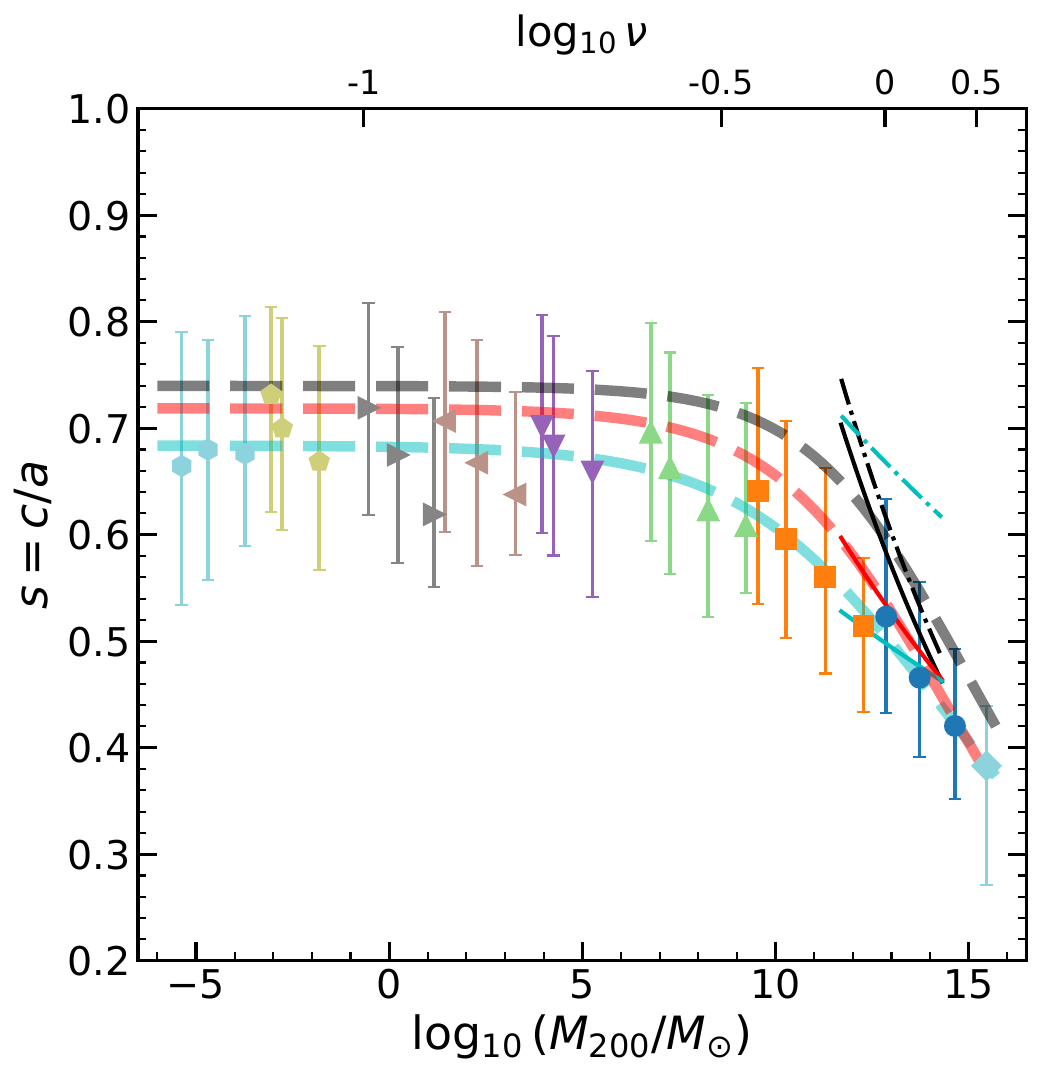}
    \includegraphics[width=0.32\textwidth]{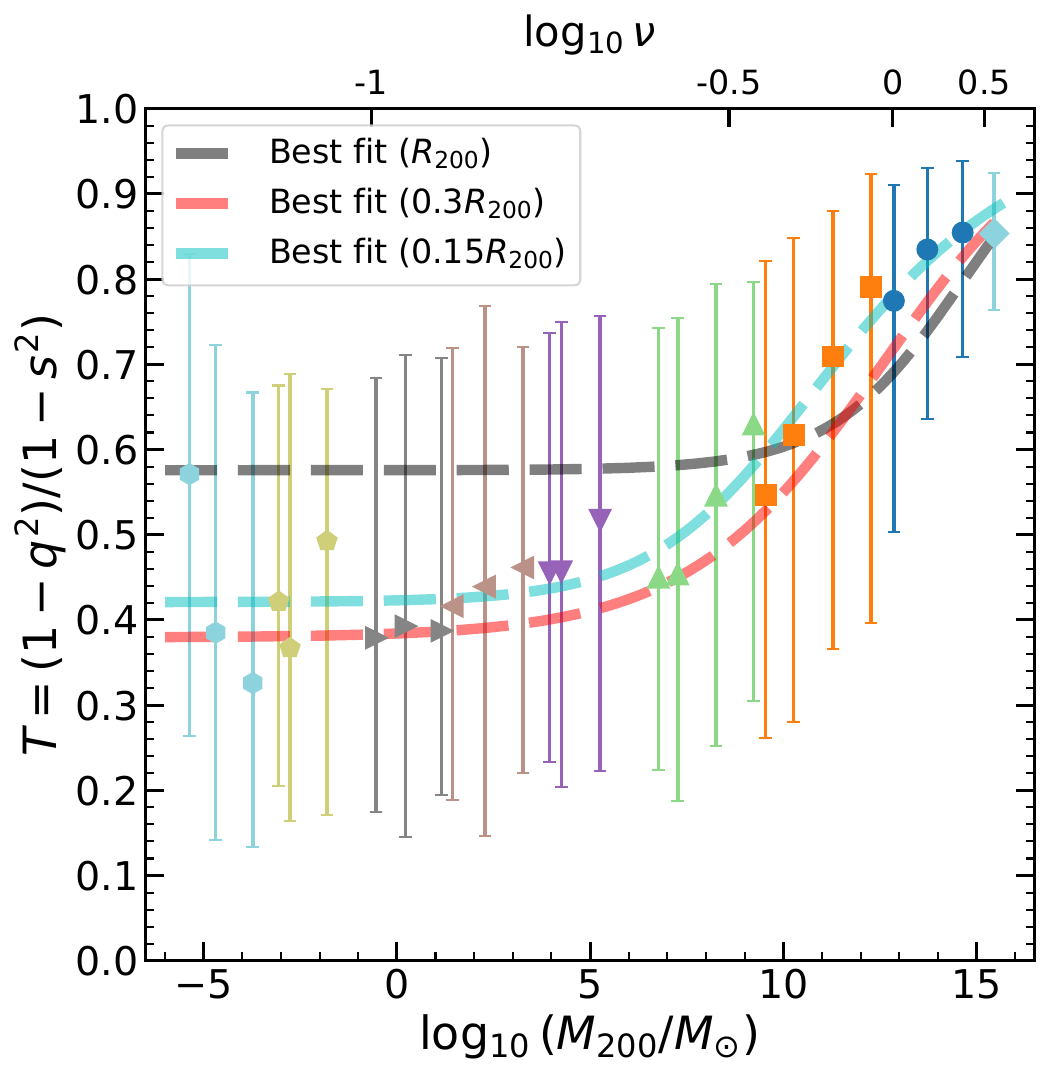}
    \caption{Mass dependence of the inner dark matter halo axis ratios $q=b/a$ (left), $s=c/a$ (middle) and $T=(1-q^2)/(1-s^2)$ (right). Symbols show the medians measured at $r=0.15R_{200}$, and error bars span the 16th--84th percentiles. Colours and symbols identify the samples as in Fig.~\ref{fig:s_r200}. Thick dashed curves show the fitting relations at $R_{200}$ (grey),  $0.3R_{200}$ (red) and $0.15R_{200}$ (cyan).}
    \label{fig:s_inner}
\end{figure*}

In this section, we examine the radial dependence of halo shapes. The left panel of Fig.~\ref{fig:s_inner} shows the measurements of $q$ at $0.15R_{200}$ with the best fitting relations at $0.15R_{200}$, $0.3R_{200}$ and $R_{200}$. The radial trend in $q$ depends on halo mass. At high masses, the median $q$ is lower in the inner shells than at $R_{200}$, consistent with previous studies \citep[e.g.][]{Chua2019}. However, at low masses, this trend appears to reverse, with slightly higher median values of $q$ in the inner shells than in the shell at $R_{200}$.

As shown in middle panel of Fig.~\ref{fig:s_inner}, the $s$ decreases towards smaller radii over all the mass range. At low masses, the fitted asymptotic values are approximately $s\sim0.69$ at $0.15R_{200}$ and $s\sim0.72$ at $0.3R_{200}$, compared with $s\sim0.74$ at $R_{200}$. The inner $s$ parameter of low mass halo therefore retain a broad flattening, but at lower value.

Previous dark matter only simulations found that massive halos tend to be more prolate in their inner regions and more oblate towards their outskirts \citep[e.g.][]{VeraCiro2011,Chua2019}. We confirm this radial ordering in our high mass haloes sample. For example, the median $T$ increases from $0.69$ at $R_{200}$ to $0.77$ at $0.15R_{200}$ for halos with $M_{200}\sim10^{13}M_{\odot}$. However, the ordering reverses in low mass range. The fitted low mass limits are $T\simeq0.42$ at $0.15R_{200}$ and $T\simeq0.57$ at $R_{200}$ indicating that the inner regions tend to be more oblate than the outskirts for low mass halos. 

Baryonic effects are more pronounced in the inner halo than in the outskirts. As the Fig.~\ref{fig:s_inner} shown, the inner dark matter distribution is substantially rounder in the full physics simulations than in their DMO counterparts and the magnitude of this change may depend on the adopted model for baryonic physics. The DMO relations presented here therefore provide a baseline for inner halo shapes, which may differ quantitatively from those of real halos. Even in low mass haloes that do not form stars, hydrodynamic effects and radiative feedback may alter the inner matter distribution \cite[e.g.][]{2024MNRAS.532.3151Z}. 

\subsection{The dynamical states of halo}

We examine the dynamical states of the halo population using the virial energy ratio and the mass central offset. These diagnostics provide complementary measures of halo relaxation \citep[e.g.][]{Neto2007,Power2012}. For virial energy ratio, we consider both the uncorrected virial ratio $\eta$ and its surface pressure corrected counterpart $\eta_{\rm corr}$.

\begin{figure}
    \includegraphics[width=\linewidth]{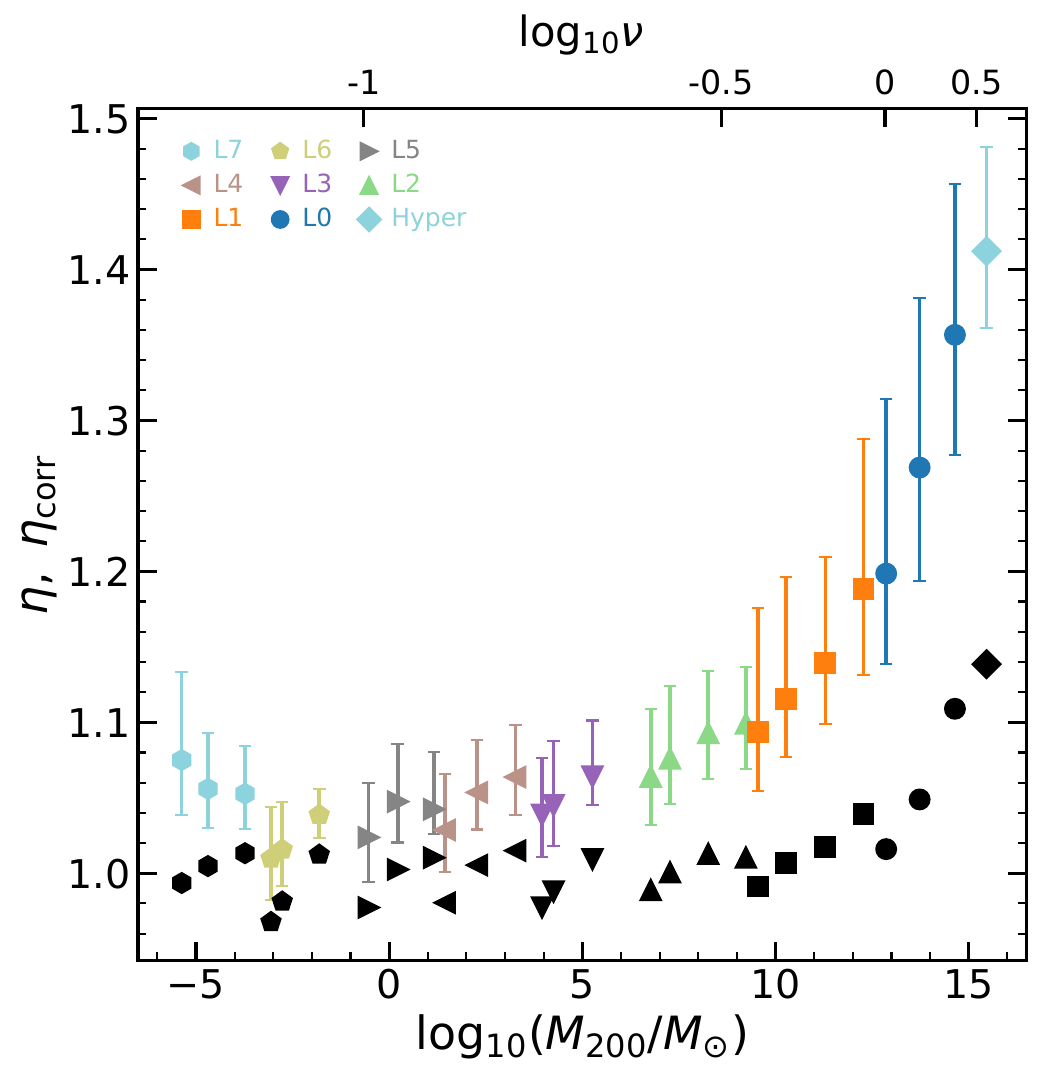}
    \centering
    \caption{Virial ratio as a function of halo mass at $z=0$. Coloured symbols show the median uncorrected virial ratio, $\eta=2K/|U|$, in each mass bin, with error bars spanning the 16th--84th percentiles. Black symbols show the median surface pressure corrected virial ratio, $\eta_{\rm corr}=(2K-E_{\rm s})/|U|$.}
    \label{fig:eta_mass}
\end{figure}

Fig.~\ref{fig:eta_mass} shows the mass dependence of the two virial ratios. The uncorrected median lies close to, but generally above, unity at low masses and increases towards the high mass end. The median $\eta$ is $1.198$, $1.269$, and $1.357$ in the bins centred at
$M_{200}=10^{13}, 10^{14}$ and $10^{15}M_{\odot}$, respectively. The median $\eta$ is $1.412$ in the most massive $10^{15.5}M_{\odot}$ bin.

As the black symbols shown, including the surface term weakens this mass dependence. The corrected median $\eta_{\rm corr}$ values for the four mass bins are $1.016$, $1.049$, $1.109$, and $1.138$, respectively. Thus, the surface term accounts for much of the excess in the uncorrected ratio $\eta$, although an excess in $\eta_{\rm corr}$ above unity remains at the highest mass halos. Accounting for the surface term makes the $\eta_{\rm corr}$ a more comparable diagnostic of dynamical relaxation across halo masses.

\begin{figure}
    \includegraphics[width=\linewidth]{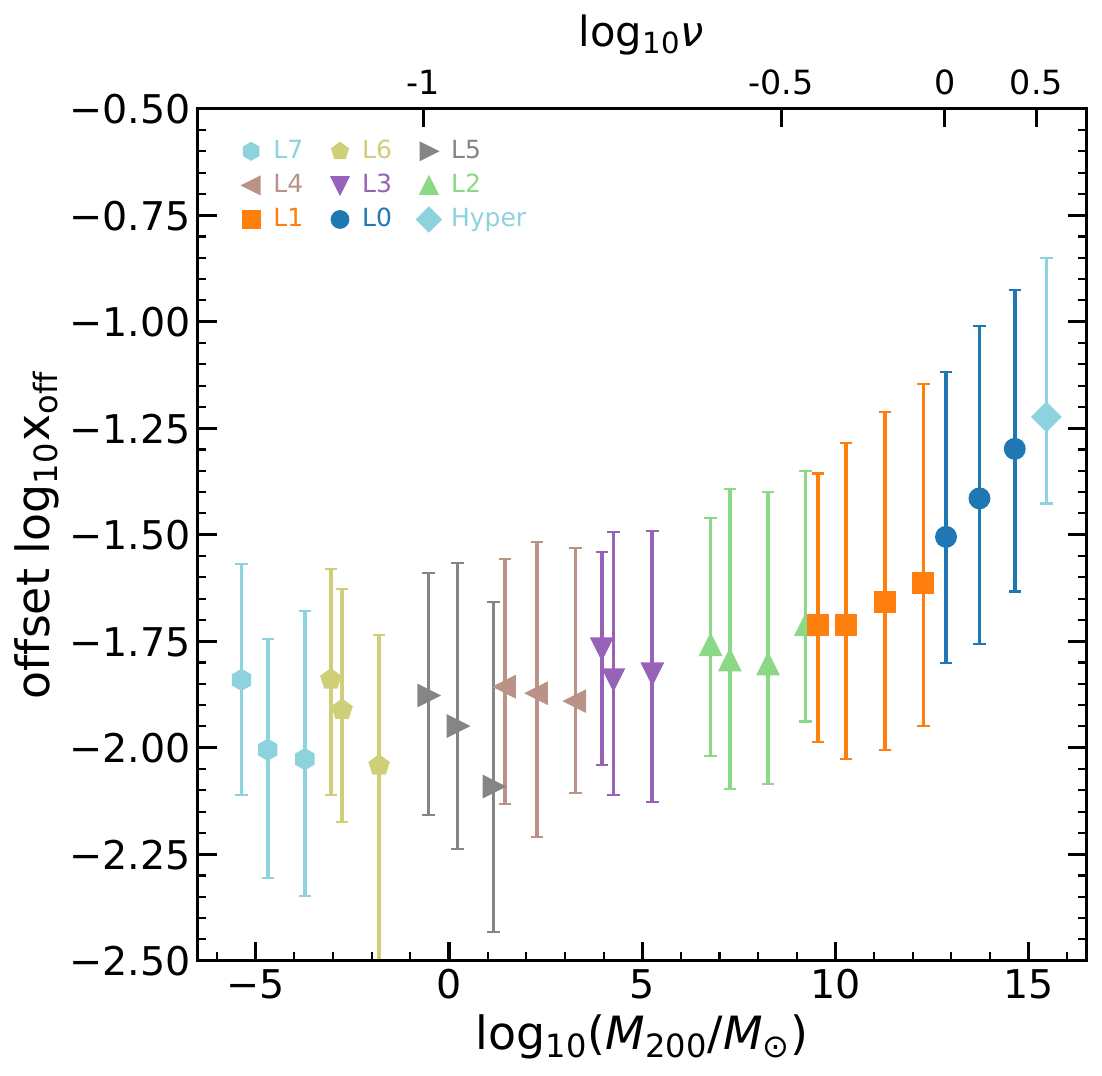}
    \centering
    \caption{Centre-of-mass offset as a function of halo mass at $z=0$. The ordinate is $\log_{10}X_{\rm off}$, where $X_{\rm off}=|{r}_{\rm cm}-{r}_{\rm c}|/R_{200}$ is the separation between the halo centre of mass and the potential minimum, normalised by $R_{200}$. Points show the median logarithmic offset in each mass bin, with error bars spanning the 16th--84th percentiles.}
    \label{fig:offset_mass}
\end{figure}

Fig.~\ref{fig:offset_mass} presents the corresponding mass dependence of offset $X_{\rm off}$. Below $M_{200} < 10^{12}{\rm M}_{\odot}$, the bin medians lie approximately between $0.008$ and $0.022$. The median $X_{\rm off}$ is $0.031$, $0.038$, $0.050$ and $0.060$ in the bins centred at $M_{200}=10^{13}, 10^{14}$, $10^{15}$ and $10^{15.5}M_{\odot}$, respectively.

Taken together, the increasing offsets and the residual excess in virial ratio are consistent with a greater departure from dynamical relaxation at the high mass. Similar with the median halo axis ratios, both dynamical state indicators also exhibit a broad plateau at low halo masses. 

\begin{figure}
    \includegraphics[width=\linewidth]{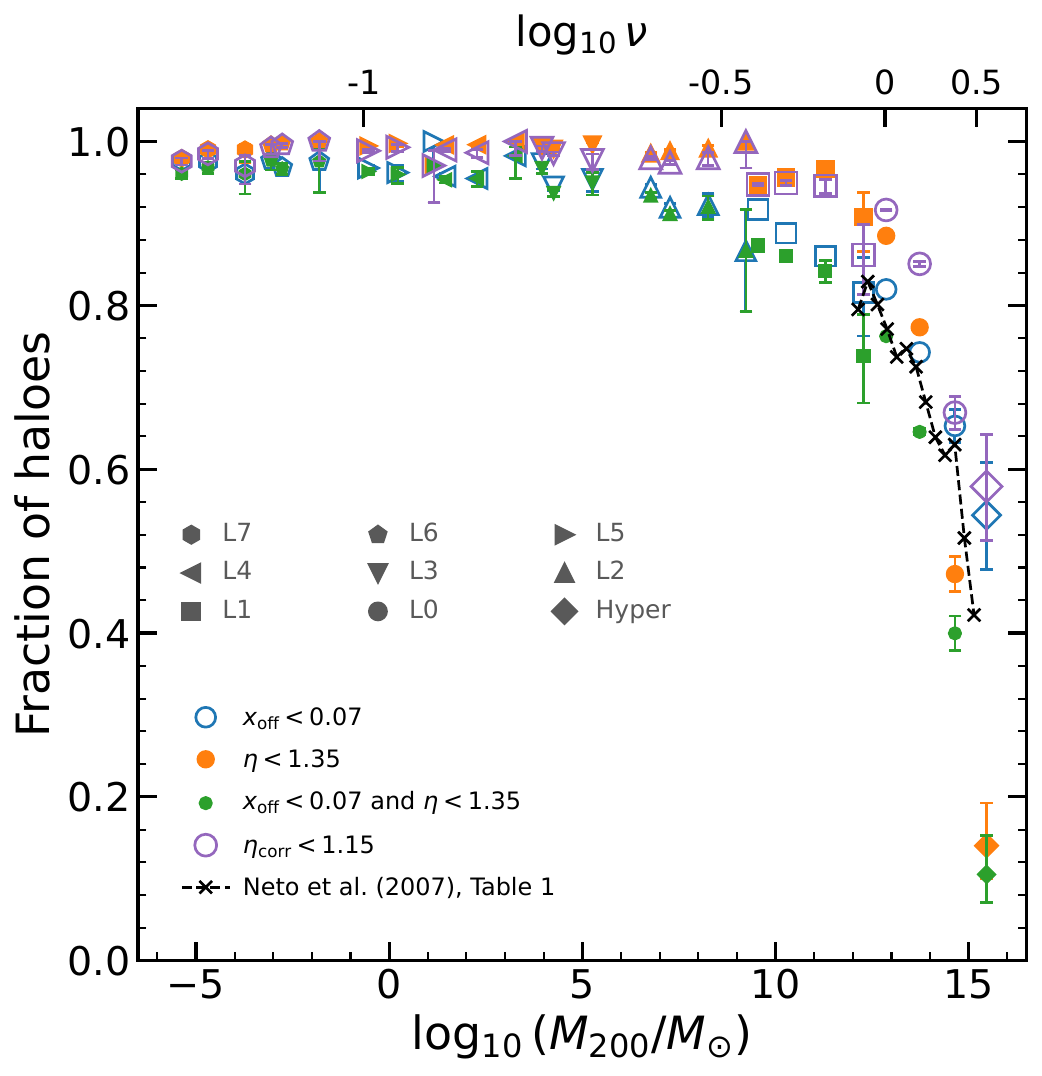}
    \centering
    \caption{Fraction of halos satisfying the adopted dynamical state criteria as a function of halo mass at $z=0$. Blue symbols show the fraction with $x_{\rm off}<0.07$, orange symbols show the fraction with $\eta<1.35$, and green symbols show the fraction satisfying both conditions simultaneously. Purple symbols require only $\eta_{\rm corr}<1.15$. Error bars show nominal $68.3\%$ Wilson binomial confidence intervals. }
    \label{fig:relaxed_fraction}
\end{figure}

Fig.~\ref{fig:relaxed_fraction} shows the fraction of halos classified as relaxed as a function of $M_{200}$ under different selection criteria. The relaxed fraction remains close to unity over a broad range of low masses and declines towards higher masses for all criteria considered. Over the overlapping mass range, our relaxed fractions are broadly consistent with those reported by \citet{Neto2007}.

\begin{figure}
    \includegraphics[width=\linewidth]{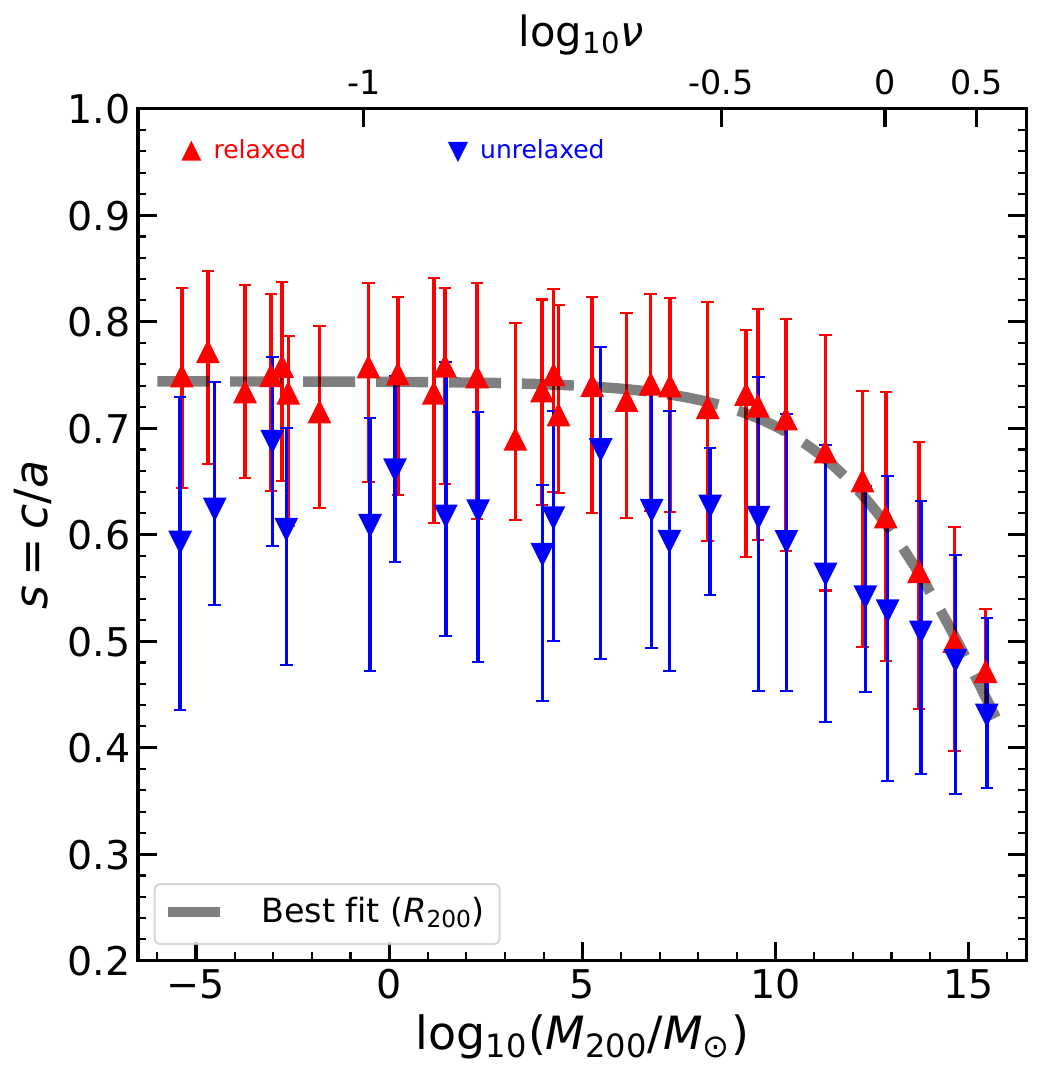}
    \centering
    \caption{Halo sphericity $s=c/a$ measured at $R_{200}$ as a function of halo mass for relaxed ($\eta<1.35$ and $X_{\rm off} < 0.07$, red upward triangles) and unrelaxed ($\eta>1.35$ or $X_{\rm off} > 0.07$, blue downward triangles) haloes. Symbols show the median $s$ in each mass bin, with error bars spanning the 16th to 84th percentiles. The grey dashed curve shows the best fitting relation for the full sample. Some mass bins are omitted due to insufficient halo numbers.}
    \label{fig:relaxed_relation}
\end{figure}

\begin{figure*}
    \includegraphics[width=0.5\textwidth]{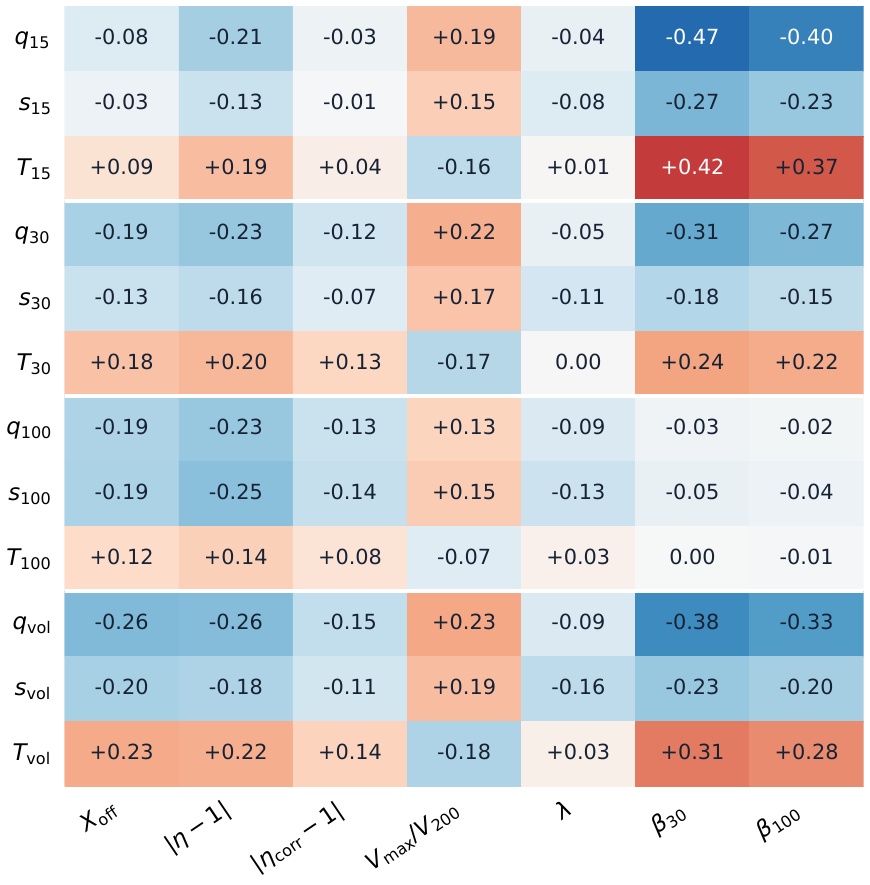}
    \includegraphics[width=0.5\textwidth]{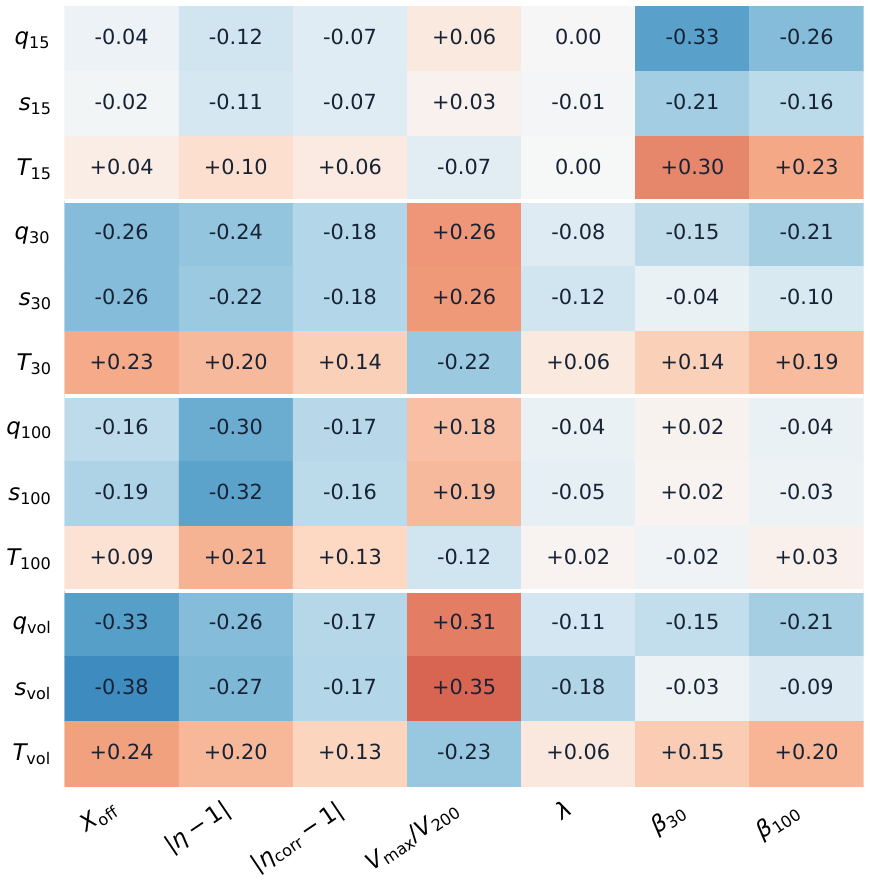}
    \caption{Correlations between halo shape and other halo properties in the low mass ($M_{200}<10^{10}\ M_\odot$, left panel) and high mass ($M_{200}\geq10^{10}\ M_\odot$, right panel) regimes. For each pair of properties, we compute the partial Spearman rank correlation within each mass bin, controlling for halo mass. The subscripts 15, 30, and 100 on the shape parameters or $\beta$ parameters denote shell measurements at $0.15R_{200}$, $0.3R_{200}$, and $R_{200}$, respectively, while $q_{\rm vol}$ and $s_{\rm vol}$ are the axis ratios measured within the ellipsoidal volume. The virial diagnostics are expressed as absolute deviations from unity $|\eta-1|$ and $|\eta_{\rm corr}-1|$. Red and blue indicate positive and negative correlations, respectively.}
    \label{fig:correlation}
\end{figure*}

\subsection{Correlations of halo shapes with halo properties}

As the previous section, we classify halos as relaxed if they satisfy both $\eta<1.35$ and $X_{\rm off}<0.07$. Fig.~\ref{fig:relaxed_relation} compares the median sphericity--mass relations of relaxed and unrelaxed halos at $R_{200}$. At low masses, the median sphericity of relaxed halos remains close to the full sample plateau at $s\sim0.74$. Unrelaxed halos are typically less spherical at comparable masses, with median values of $s\sim0.6$ in most low mass bins. Thus, even among low mass halos that satisfy relaxation criteria, the typical halo shape departs appreciably from spherical symmetry. Notably, halos can retain obvious triaxial shapes even when they are close to dynamical equilibrium \citep[e.g.][]{2004MNRAS.354..522M}.

The median sphericity decreases towards the high mass end in both populations and the difference between their medians narrows in the massive halo bins. Hence, the overall trend in halo shape reflects both the increasing fraction of unrelaxed halos and the changes in halo shape among relaxed halos as mass increases. With this definition of relaxation, the small difference in median sphericity at the high mass regime is also consistent with previous works \citep[e.g.][]{Henson2017}.

The scatter of sphericity remain broad and overlap substantially within individual mass bins in Fig.~\ref{fig:relaxed_relation}. Halo mass and relaxation class therefore capture only part of the halo variation in halo shape.

Fig.~\ref{fig:correlation} shows the correlations between halo shapes and various halo properties. For each pair of variables, we compute the partial Spearman rank correlation separately in each mass bin to remove the dependence of each variable on halo mass. Then, each matrix entry gives the unweighted average of these coefficients across the bins in the corresponding mass range. We also examined the median coefficients, finding that they differ only slightly from the mean values. Although overall trends and correlation signs remain broadly consistent, their strengths vary with halo radius and mass. 

In both mass regimes, inner halo shapes exhibit a clear association with velocity dispersion anisotropy. At $0.15R_{200}$, the bin averaged partial Spearman coefficients between $q_{15}$ and $\beta_{30}$ are $-0.47$ and $-0.33$ for the low and high mass samples, respectively, while the corresponding coefficients for $T_{15}$ are $+0.42$ and $+0.30$. Thus, more radially biased velocity distributions are associated with less spherical and more prolate halo configurations. These associations are stronger in the low mass population and weaken substantially towards $R_{200}$, where the shape--$\beta$ correlations are close to zero. 

The axis ratios generally exhibit negative correlations with $X_{\rm off}$ and the absolute deviations of the $\eta$ from unity. These results show a statistical association between the dynamical states and the non-sphericity of the halo mass distribution. The correlations are consistently weaker when the surface pressure correction is included. It is consistent with previous work which show the surface correction can complicate the relationship between the virial energy ratio and assembly activity \citep[e.g.][]{Power2012}.

The $V_{\max}/V_{200}$ correlates positively with the axis ratios, most clearly for the global shape at high masses, where the coefficient for $s_{\rm vol}$ reaches $+0.35$. Interestingly, in the high mass sample, the correlations with $V_{\max}/V_{200}$ are much weaker at $0.15R_{200}$, in agreement with the weak correlations between inner halo shape and concentration \citep[][]{Chua2019}. At fixed mass, this correlation suggests that halos with more centrally concentrated mass distributions tend to have rounder shapes. The association may partly reflect a shared dependence on assembly history, since earlier forming halos tend to be both more concentrated and more spherical \citep[e.g.][]{2006MNRAS.367.1781A}.

Correlations with halo spin $\lambda$ are comparatively weak, with absolute mean coefficients below approximately $0.2$ throughout. This pattern is consistent with the picture that triaxial halo shapes are sustained primarily by anisotropic velocity dispersion rather than by rotation \citep[e.g.][]{1996MNRAS.281..716C, 2005MNRAS.363..937B}.

\section{Summary and Discussion} \label{sec:res}

Taking advantage of VVV and HM simulations, we have investigated the shapes and dynamical states of dark matter halos across the full mass spectrum in a standard $\Lambda$CDM. Our sample spans more than twenty orders of magnitude in mass, from the most massive cluster halo with $M_{200}\sim10^{15.5}M_{\odot}$ down to smallest planetary mass halos with $M_{200}\sim10^{-6}M_{\odot}$. Our main findings are as follows.

\begin{enumerate}

\item More massive halos are systematically less spherical and more prolate, while the mass dependence of halo shapes becomes progressively weaker towards lower masses. The median axis ratios are well described by smoothly broken power laws, approaching low mass plateaus of $s\sim0.74$ and $q\sim0.87$, and decreasing towards $s\sim0.42$ and $q\sim0.56$ at the high mass end. Correspondingly, the median triaxiality parameter increases from $T\sim0.57$ over most of the low-mass regime to $T\sim0.86$ for the most massive halos.

\item The radial dependence of halo shape changes with mass. The inner halo regions have smaller median $s$ than with the low mass plateau of $s\sim0.69$ at $0.15R_{200}$, $s\sim0.72$ at $0.3R_{200}$ and $s\sim0.74$ at $R_{200}$. The radial ordering of $q$ and $T$ reverses between the two mass regimes. At low masses, the inner shells tend to have larger $q$ and lower $T$ than the outskirts, whereas the inner shapes of massive halos tend to have smaller $q$ and higher $T$. The best-fitting parameters for the shapes--mass relations are listed in Table~\ref{tab:fit_params}.

\item  The median virial ratio $\eta$ and offset $X_{\rm off}$ vary weakly with mass at the low mass regime and increase towards cluster masses. Accounting for the surface pressure term reduces the mass dependence of the virial ratio. The fraction of relaxed halos satisfying $\eta<1.35$ and $X_{\rm off}<0.07$ is close to unity over a broad range of low masses and drops sharply towards the high mass end.

\item The similar low mass shapes plateau persists among halos that satisfy relaxation criteria. Unrelaxed halos are typically less spherical at comparable low masses. The difference between the two populations narrows in the most massive bins, although both two populations become less spherical towards higher masses. Because unrelaxed halos make up only a small fraction of the low mass population, the median shape--mass relation for relaxed sample closely follows that of the full sample.

\item After removing the mass dependence of halo shapes and dynamical properties, we find that residual variations in halo shape correlate with the internal dynamical states. We find that more radially biased velocities are associated with smaller axis ratios and more prolate shapes. Halo shapes also show moderate correlations with the concentration proxy $V_{\rm max}/V_{200}$, with higher values associated with rounder shapes, but only weak correlations with halo spin.
\end{enumerate}

As reported by previous works \citep[e.g.][]{2006MNRAS.367.1781A}, earlier forming halos tend to be more spherical. One possible explanation for the flattening of the shape--mass relation is that halo formation times depend less strongly on mass in the low mass regime \citep[e.g.][]{2024MNRAS.52711740L}. 


Furthermore, previous simulations suggest that the redshift evolution of median halo shapes is relatively weak at fixed peak height \citep[e.g.][]{2014MNRAS.443.3208D, Bonamigo2015, 2017MNRAS.467.3226V}. If this behaviour also holds over the mass range explored here, our results may provide a useful reference for estimating typical halo shapes at higher redshifts by matching halo peak heights to those indicated by the upper axes of our figures. We leave the quantitative test of this extrapolation to future work.


\section{ACKNOWLEDGMENTS}
\begin{acknowledgments}
HY thanks Haonan Zheng and Shihong Liao for thorough discussions and comments. This work is supported by the National Natural Science Foundation of China (Grant No. 12588202) and the National Key Research and Development Program of China (Grant
No. 2023YFB3002500). QG acknowledges the support from the National Natural Science Foundation of China (Nos. 12425303). ML also acknowledges the support from the National SKA Program of China (Nos. 2022SKA0110201) and the National Key Research and Development Program of China (No. 2022YFA1602903). QW also acknowledges the support from National Natural Science Foundation of China (NSFC) (Nos. 12473002, 12473015)
\end{acknowledgments}

\vspace{5mm}





\appendix
\section{Effect of substructures in shape determination}
\label{sec:A1}

\begin{figure*}
    \includegraphics[width=0.32\textwidth]{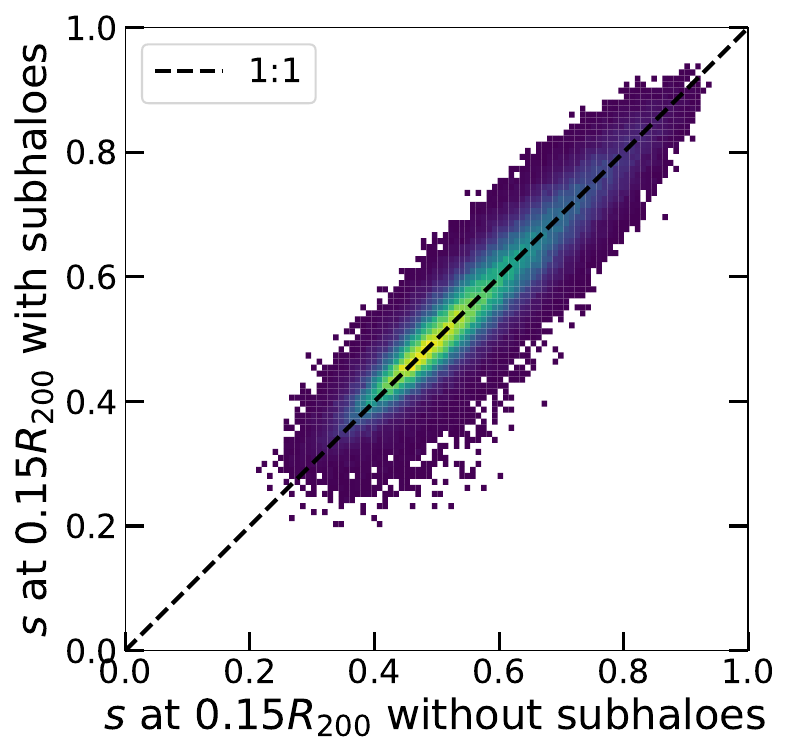}
    \includegraphics[width=0.32\textwidth]{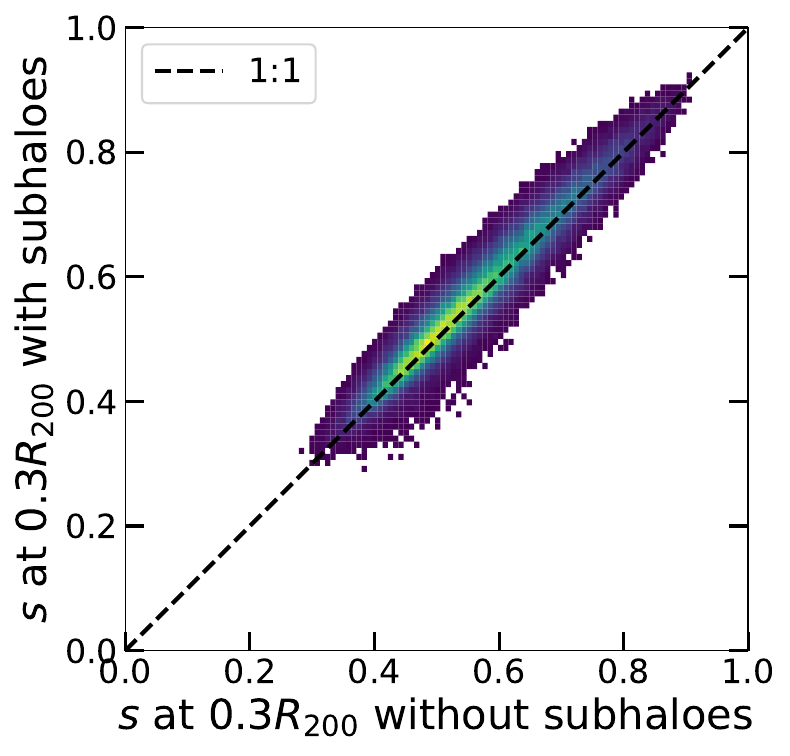}
    \includegraphics[width=0.32\textwidth]{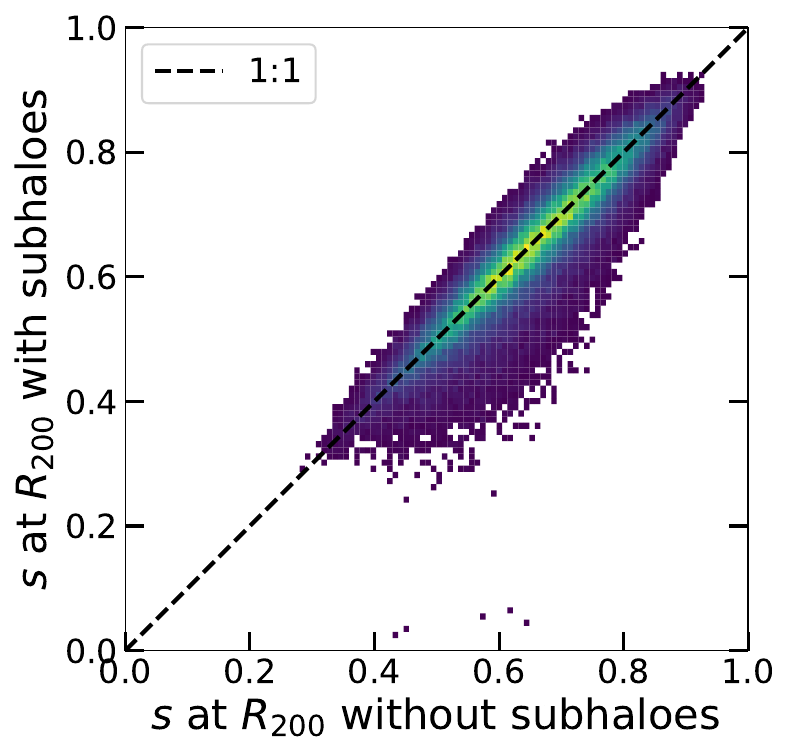}
    \caption{Effect of substructure on the local $s=c/a$. The panels show measurements in ellipsoidal shells at $0.15R_{200}$, $0.3R_{200}$, and $R_{200}$, from left to right. In each panel, the horizontal and vertical axes show $s$ measured excluding and including particles bound to subhalos for the same halos, respectively. The dashed line indicates equality. The two measurements generally agree, although including subhalo particles lowers the measured $s$ for some halos in the outer shell.}
    \label{fig:sub_effect}
\end{figure*}

Our fiducial measurements include particles bound to subhalos. To test how this choice affects the local halo shape in this work, we repeat the measurements after removing these particles and compare the axis ratio $s$ for the same halos
(Fig.~\ref{fig:sub_effect}).

The measurements generally agree at all three radii, indicating that the local axis ratio is typically insensitive to whether subhalo particles are included. The most noticeable departures occur in the outer shell, where including subhalo particles lowers the measured $s$ for some halos \citep[e.g.][]{2011ApJS..197...30Z}. Substructure can therefore
affect the inferred outer shape of individual halos even though the two measurements generally agree. The close agreement suggests that including subhalo particles is
unlikely to substantially affect our main conclusions about halo shapes. Since the contribution of substructure is difficult to
isolate in real observations, we retain particles bound to subhalos as our fiducial results.

\section{Dependence of Halo Shape on the Measurement Methods}
\label{sec:A2}

\begin{figure*}
    \includegraphics[width=0.32\textwidth]{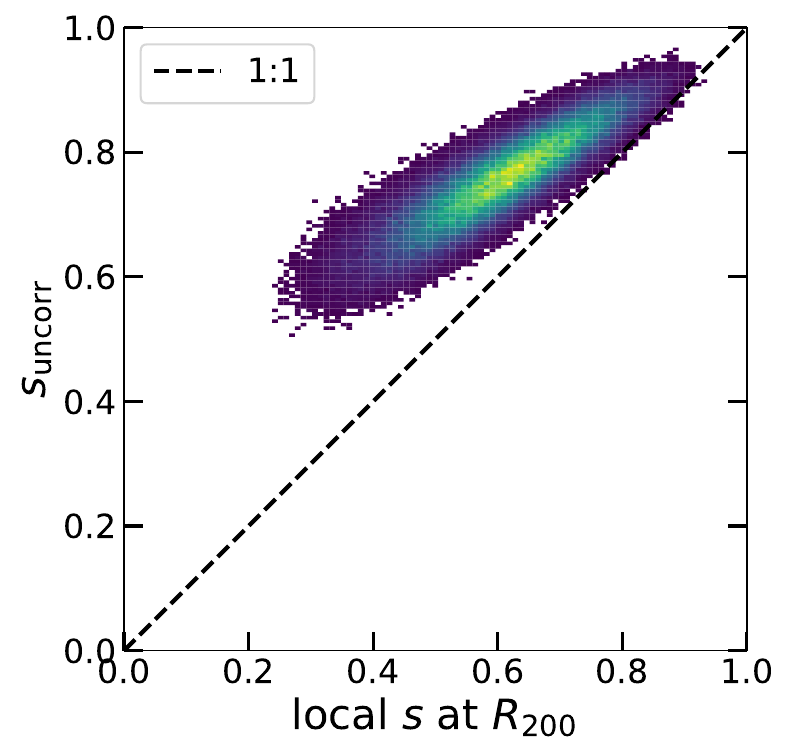}
    \includegraphics[width=0.32\textwidth]{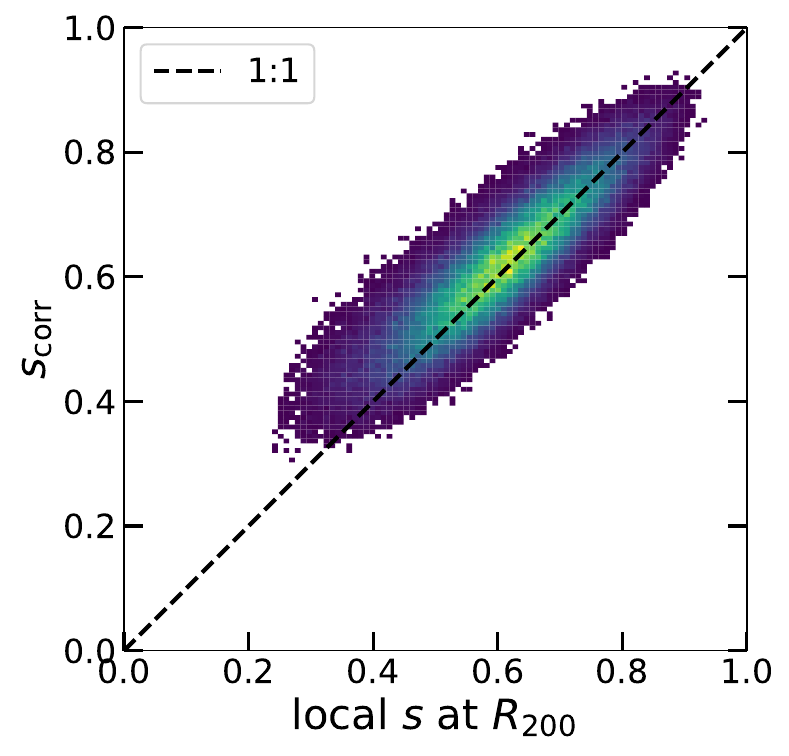}
    \includegraphics[width=0.32\textwidth]{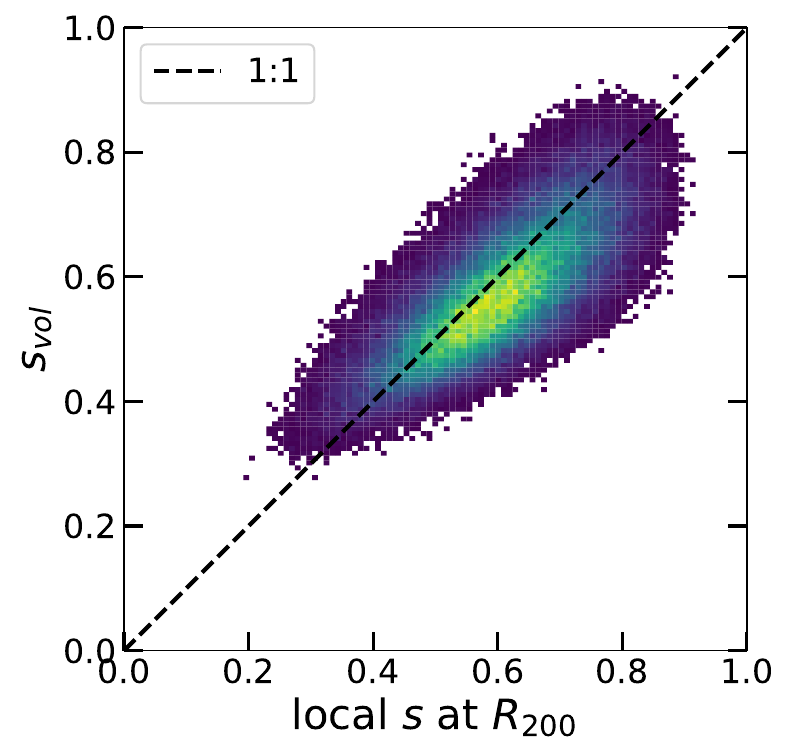}
    \caption{Comparison of the fiducial local $s=c/a$, with three estimates of the global axis ratio for the same haloes in L0. In all panels, the horizontal axis shows $s$ measured iteratively in an ellipsoidal shell at $R_{200}$. From left to right, the vertical axis shows (i) $s$ measured within a sphere of radius $R_{200}$ using the unweighted shape tensor ($w=1$); (ii) the same measurement after applying the correction $s_{\rm corr}=s_{\rm raw}^{\sqrt{3}}$; and (iii) $s$ measured within an ellipsoidal volume using the reduced shape tensor ($w=r_{\rm ell}^{2}$). The spherical measurements are performed without iteration. The dashed line indicates equality.}
    \label{fig:shape_method}
\end{figure*}

We compare the shape measurement methods using the same halos from the VVV-L0 simulation, retaining particles bound to subhalos in all measurements. As shown in Fig.~\ref{fig:shape_method}, the unweighted shape tensor evaluated within a spherical aperture tends to yield larger $s$ than our fiducial measurement in an ellipsoidal shell at $R_{200}$. A spherical aperture biases the inferred shape towards rounder values, while the unweighted tensor gives particles at larger radii greater leverage. Applying the correction $s_{\rm corr}=s_{\rm raw}^{\sqrt{3}}$ \citep{2005ApJ...627..647B} brings the axis ratios into close agreement with our fiducial measurements. The spherical aperture correction brings the axis ratios into
close agreement with our fiducial shell measurements at $R_{200}$.
The unweighted shape tensor is also useful for comparisons with weak lensing studies because its projection describes the second moments of the projected mass distribution
\citep{2012MNRAS.420.3303B, 2013MNRAS.429.3316B}. The reduced tensor evaluated over an ellipsoidal volume, by comparison, tends to yield slightly smaller $s$ than the local measurement at $R_{200}$. This is consistent with the volume measurement incorporating inner regions, where inner $s$ is typically smaller in our sample.

\section{Sensitivity of halo properties to local environment}
\label{sec:A3}

\begin{figure}
    \includegraphics[width=0.5\linewidth]{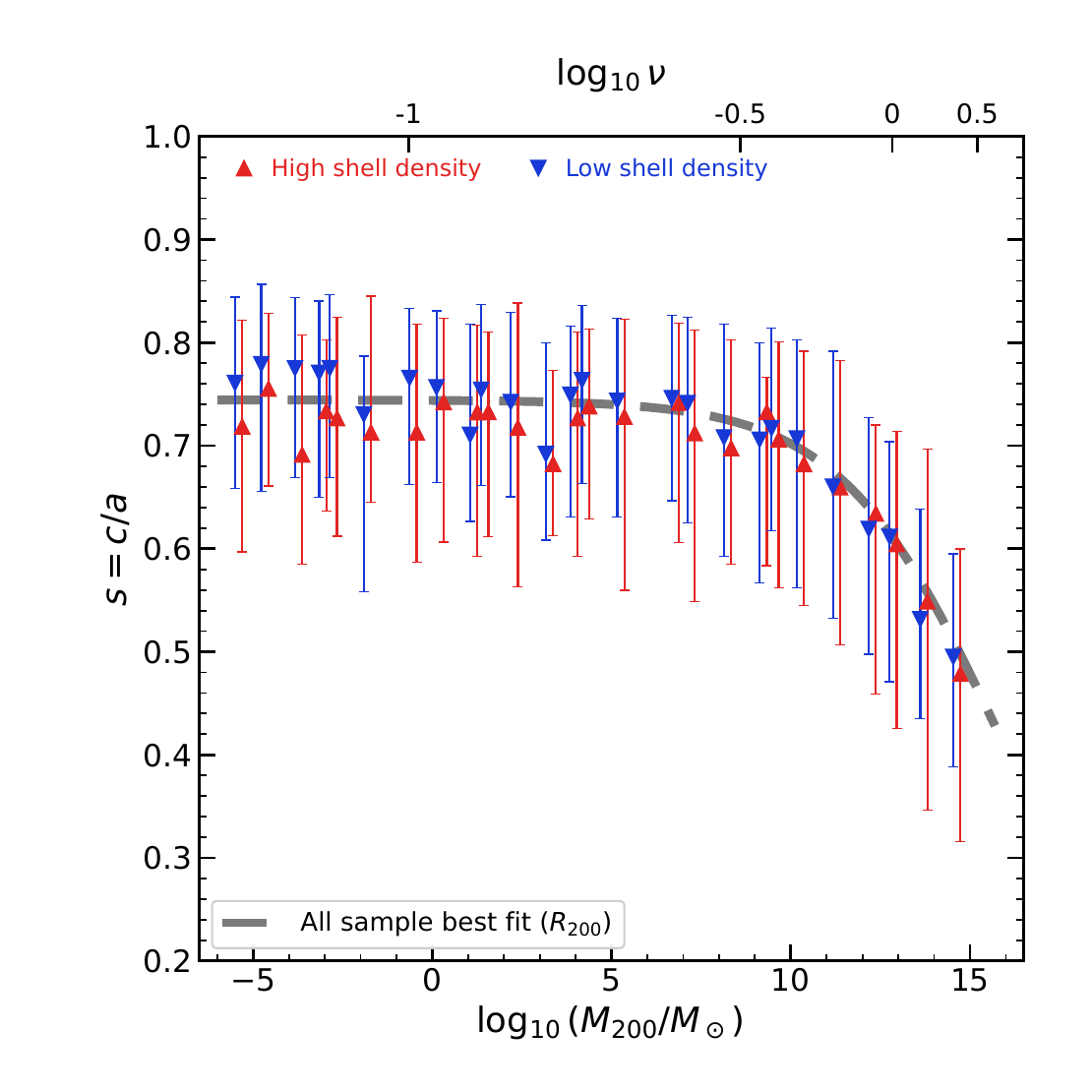}
    \centering
    \caption{Halo sphericity $s=c/a$ measured at $R_{200}$ as a function of halo mass for halos in high density (red upward triangles) and low density (blue downward triangles) local environments in VVV simulations. The two subsamples are defined by splitting the mean density in the spherical shell $2<r/R_{200}<3$ at its median within narrow mass intervals.}
    \label{fig:sub_env}
\end{figure}

The L1--L7 zooms in VVV preferentially sample underdense regions, so halos in these runs may differ from a volume representative population of the same mass. To assess the dependence of halo shapes on environment, we estimate the local environmental density within the spherical shell $2R_{200}<r<3R_{200}$ around each halo. We then divide halos into high and low density subsets at the median local density within each mass bin and compare their halo properties distributions using the KS test. We find no significant differences between the two subsets. Fig.~\ref{fig:sub_env} illustrates this comparison using $s$ as a representative shape parameter. This weak environmental density dependence is consistent with previous findings \citep{2007MNRAS.378...55M, 2007MNRAS.375..489H, 2020Natur.585...39W}.

\bibliography{sample631}{}
\bibliographystyle{aasjournal}



\end{document}